# "The Pinpoint Comets: 133P/Elst-Pizarro, 249P/LINEAR, 331P/Gibbs, 62412 and 6478 Gault"

Ignacio Ferrín
Institute of Physics, FACOM and SEAP, University of Antioquia,
Medellín, Colombia, 05001000
ignacio.ferrin@udea.edu.co

Cesar Fornari
Observatorio "Galileo Galilei", X31
Oro Verde, Argentina

Agustín Acosta
Observatorio "Costa Teguise", Z39
Lanzarote, España

**Abstract**

From two Active Asteroid (AA) known in 1979 we have advanced to 37 members at the beginning of 2019. More surprisingly, in the first three months of 2019 five new members were added to the list, one of them, 6478 Gault, with a curious, out of the ordinary tail. This prompted a visual search for similar objects in comet's images depositories. We have identified five cometary objects that have the same morphological appearance: A thin long tail coming out of a star-like nucleus with no gas. The members of this new class (Pinpoint Comet Group, PCG) are, 133P/Elst-Pizarro, 249P/LINEAR, 331P/Gibbs, 62412 and 6478 Gault. We look for differences and similarities among them. Four of them belong to the main belt and one to the Jupiter Family of comets. One of them has left the group but will return soon. Two of them are double. All exhibit densities well above the Jupiter Family of comets values. This tells us something about their origin and evolution. In this work we study the secular and rotational light curves of these objects and their phase and SLC plots. We will investigate the meaning of these characteristics in terms of their physical evolution.

## 1.1 Introduction

At the beginning of 2019 a new main belt comet (MBC) was discovered by the Asteroid Terrestrial-Impact Last Alert System, ATLAS, (Smith et al., 2019). The appearance of 6478 Gault immediately called attention because the nucleus did not exhibit a coma and the tail was very long and thin with no evidence of gas. This morphology is quite different from the usual comet that exhibits a gaussian shaped coma and dust tail that expands and fades with distance to the nucleus.

Studies of the MBCs (Jewitt, 2014; Hsieh et al., 2004; Snodgrass et al., 2017) have improved our understanding of this group significantly. However the MBCs keeps surprising us. This new member 6478 exhibits characteristics significantly different to the rest of the MBC group. In this work we will investigate the meaning of these characteristics in terms of their physical evolution.

Up to the moment of this writing there have been 8 articles related to 6478 Gault. It is appropriated to discuss them here because much of what has been learned of 6478 applies to the other members of the PCG. The last two papers are of our own production (Ferrín and Acosta, 2019a; Ferrín et al., 2019b).

Ye et al. (2019) describe two brightening events that correspond to our active zones Z1 and Z2, and that released 2x107 and 1x10(6) kg. Dust dynamics showed grains of up to 10 microns in size ejected at velocities less than 1 m/s regardless of particle size. Additionally they derive an upper limit to the ejection velocity of < 8 m/s.

Kleina et al. (2019) determine an absolute magnitude of 14.4 in the V band, and thus derive a nucleus 4 km in diameter with an adopted albedo of 0.20 representative of asteroids. They also adopt a density of 3000 kg/m$^3$. Although they have an extensive series of observations these did not show a clear light curve and the rotational period could only be estimated at ~1h for one peak, and ~2h for two peaks. They derived a $v_{eject}$ = 0.7 m/s for the maximum emission velocity.

Moreno et al. (2019) adopt for the dust density 3400 kg/m$^3$ which is appropriated for S-type asteroids. They also adopt a geometric albedo pv=0.15 typical of S-types. With an absolute magnitude Hv= 14.4, they find a diameter of 4.5 km. They also apply a phase coefficient of 0.033

mag/°. Long series of observations secured with the TRAPPIST-North and South did no exhibit evidence of a rotational signature.

Jewitt et al. (2019) find mass loss rates of ~10-20 kg/s and a typical particle radius in the main tail <a> ~ 1000 μm. Using the thickness of the tail at the time of orbital plane crossing they are able to calculate a value for the ejection velocity $v_{eject}$ = ~ 2 m/s. They also find a close similarity to comet 311P/2013 P5 who exhibited multiple dust ejections (Jewitt et al., 2018).

Hui et al. (2019) confirm the slow dust ejection velocity for the dust, finding $v_{eject}$ = 0.15 m/s. As previous authors they infer that the mass-loss events were caused by rotational instability.

Chandler et al. (2019) used archival observation to demonstrate that Gault has a long history of previous activity. Their data suggests that activity is caused by a body spinning near the rotational limit. They recognize 6478 as a new class of object perpetually active due to rotational spin up. Their data is incorporated on our plots.

It is important to point out that none of the six manuscripts published on this object have a value for the rotational period, thus their conclusions on the reason of activity is entirely theoretical. However their prediction is correct since this work finds a rotational period Prot = 3.360±0.005 h. We find that the activity is not perpetual but episodic. Additionally, according to our data, another comet that resembles 6478 is 133P/Elst-Pizarro as shown later on in Section 2.1.

All the above works adopted for their calculations an absolute magnitude Hv = 14.40 listed by the JPL site. The MPC lists Hv = 14.30. Our own observations studied later on (Section 6) will show an absolute magnitude $m_V(1,1,a)$ = 16.12±0.05. Consequently some of the calculations performed in the above works might have to be revised.

Additional works might be relevant to this investigation.

The conclusion by Hshie et al. (2004) that *"apparent low ejection velocities of optical dominant dust particles, as implied by the lack of an observable sunward dust extension or coma-like dust halo around the nucleus, and the narrow width of the dust trail, make extremely small particles unlikely to be optically significant".* This result is consistent with the hypothesis that these large particles are the remnant of the cometary activity deposited on the surface. It is also consistent with the idea that these objects might be completely covered with a thick mantle of dust.

Kokotanekova et al. (2017) studied an ensemble of Jupiter-family comets, finding a cut-off in bulk densities at 0.6 g/cm3 if the objects are strengthless. They also find an increasing linear phase function coefficient with increasing albedo. And the median linear phase function coefficient for JFCs was 0.046 mag/deg and the median geometric albedo was 4.2%. These values are useful when this information is not available for a given object.

Hsieh et al. (2009) note that low albedos ($p_R$ < 0.075) remains a consistent feature of all cometary (i.e., icy) bodies, whether they originate in the inner solar system (the MBCs) or in the outer solar system (all other comets).

A resume of our own results on 6478 Gault will be presented in Section 6.2.

**1.2 - Definition of Pinpoint comet**

A pinpoint comet is a comet-like object complying with the following conditions:

(1) a star-like nucleus with no gas coma

(2) no gas tail

(3) thin, long dust tail; in fact the ratio length/width > 50 is characteristic of the PCG, while normal comets have length/width ≤ 10. Comet 249P/LINEAR exhibited a length/width > 80 in original images.

(4) tail does not widen with distance or widens very slowly. Jewitt et al. (2019) measured the width as a function of distance to the nucleus to derive an expansion velocity that they found very small.

(5) sustained activity over time (months, years). This condition is needed to distinguish this group from comets exhibiting trails which also show thin and long tails but for a short duration of time (days), while Earth crosses the comet´s orbital plane.

(6) Theoretical interpretations of 6478 Gault (Chandler et al., 2019; Hui et al., 2019; Jewitt et al., 2019; Kleyna et al., 2019; Moreno et al., 2019; Ye et al., 2019), agree that this object is in a state of rotational disruption. This state should also be applicable to the other members of this complex, which imposes another condition, a rotational period in the range of 3 to 4h. Curiously the only three periods we know (3.33 h, 3.36 h, 3.48 h), show very small dispersion (ΔProt = ±0.05 h).

Figure 1 shows images of the 5 members of the PCG that have been identified up to now, although one member, 249P, has left the group for a short period of time, as we will see later on.

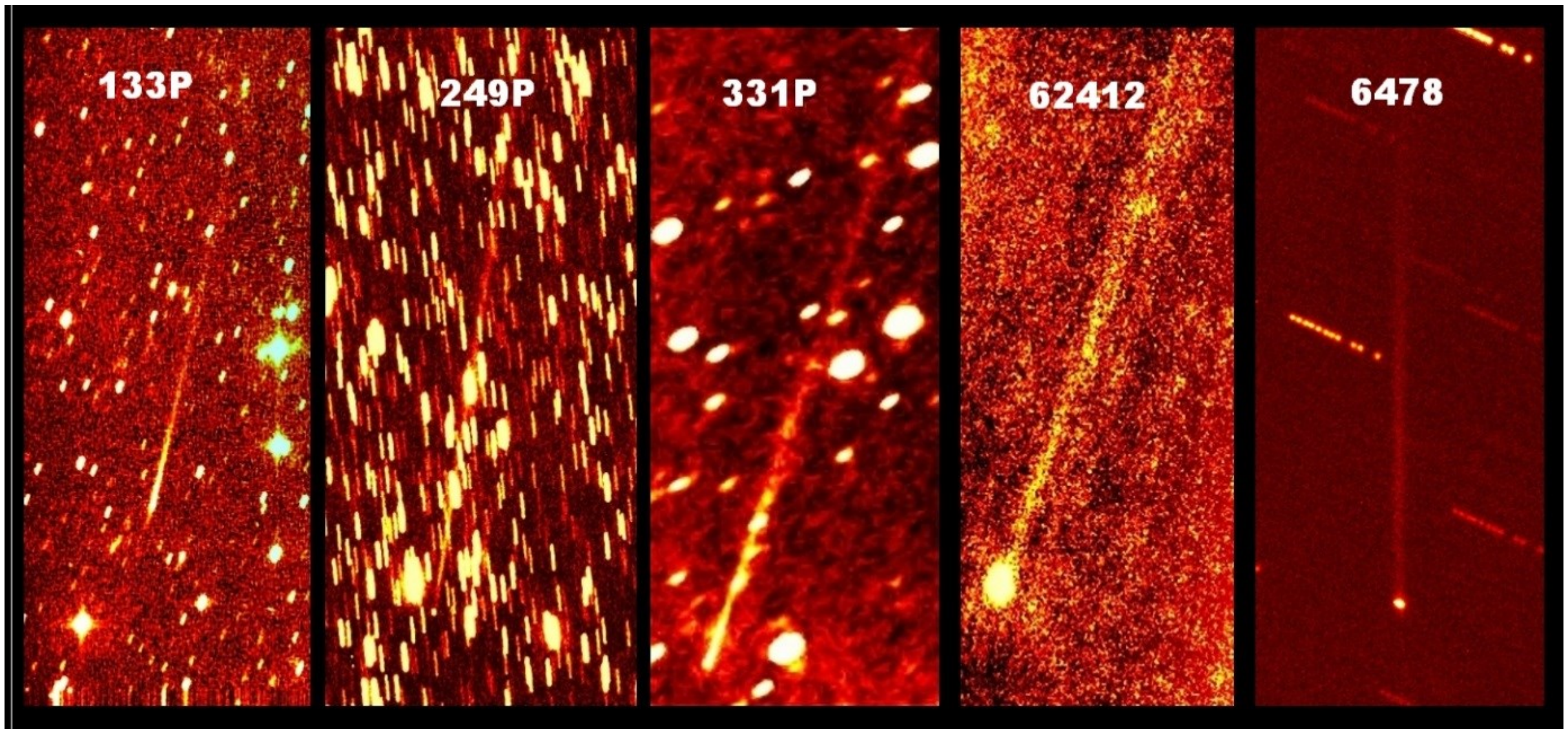

**Figure 1.** Images of the 5 currently known members of the Pinpoint Comet Group (PCG). All of them exhibit a star-like head with no coma, a persistent, thin, long, gas-less tail that does not open with distance, characteristics of this group. From left to right: (133P/Elst-Pizarro) Jim Scotti, (249P/LINEAR) Michael Jeager and Gerald Rhemann. (331P/Gibbs) José Francisco Hernandez. (62412) Scott Sheppard and Chadwick Trujillo. (6478/Gault) Cesar Fornari. The ratio length/width ~55, ~92, ~68, ~51, ~63 respectively, is remarkable, especially for 249P. The image of 62412 is the only one available in the literature. All images are reproduced with permission.

<> <> <> <> <>

We would like to answer several questions: What do these objects have in common? In what do they differ? Are they sublimating or are they suffocated? Do they represent a terminal stage in their lifetime? Why did 249P left the group after 2006? Is there evidence of duplicity in any of them? May their density be considered normal? Are these traits restricted to main belt comets? To try to answer these questions we will make use of the secular light curve formalism (SLCs, Ferrín, 2010). The SLC formalism depicts the absolute magnitude of the object versus time from perihelion, from aphelion (-Q) to aphelion(+Q) giving the whole orbital picture.

In Section 2, we will study 133P/Elst-Pizarro. Sections 3, 4, 5 and 6 will be dedicated to 249P/LINEAR, 331P/Gibbs, 62413 and 6478Gault, respectively. In Section 7 we will consider the evolution of the PCG and in Section 8 we will present the agreement between theory and evolutionary plot, and in Section 9 we list a resume of our results.

### 1.3 Quality of our photometry

We carried out the photometry of 62412 using several telescopes listed in Table 1. The photometry of 6478 Gault was presented previously (Ferrín and Acosta, 2019a, Ferrín et al., 2019b). The photometry of 62412 is listed in Table 2, while Table 3 and 4 give respectively the parameters determined in this work and the rotational periods of the members of the PCG. Table 5 lists the orbital elements of the PCG and their Tisserand parameter values.

We are confident that our photometry is of high quality because we have adopted procedures to that effect (Ferrín et al., 2019b). Images were processed with darks and flats in the usual manner. We adopted large photometric apertures typically 4-5 FWHM in order to extract the whole flux from the objects. We used from 5-15 comparison stars, and rejected those that were outliners. We used APASS, the best V-magnitude catalog that spans the whole sky, created by the AAVSO to cover the observation of variable stars. We measured the observed magnitude from long series of images, typically containing 10-50 images with 3-5 minutes of exposure time each, reaching to total exposure times of hours, to plot one single photometric data point per night. Before combining, each image was examined for defects, cosmic rays, clouds, star contamination, disappearance of the object, reflections and bright stars. Unsuitable images were discarded before combining. The combination algorithm was the median, to clean up cosmic rays and defects. Finally, we adopted the envelope of the data points as the correct interpretation of the light curve. The envelope of the data points represents a perfect instrument+CCD, with a perfect atmosphere, capable of extracting the whole flux from the object and thus the maximum relative brightness (Ferrín, 2005).

As a result we routinely achieve an error of $\pm 0.01$ to $\pm 0.02$ magnitudes equivalent to 1-2% error. The error is so small that it is always contained inside the plotting symbol.

### 2.1 133P Elst-Pizarro: Introduction

On 1979 the first two members of the Activated Asteroids Complex (MBCs) were discovered: 107P/Wilson-Harrington and 133P/Elst-Pizarro. The first one exhibited a very feeble activity (Ferrín, 2012) while the second was discovered later on exhibiting a long thin tail quite different from the usual comet tails, as can be seen in Figure 2 and it turned out to be the first member of the PCG (Hsieh et al., 2004). Additional images can be found at the following link: http://www.aerith.net/comet/catalog/0133P/2013-pictures.html

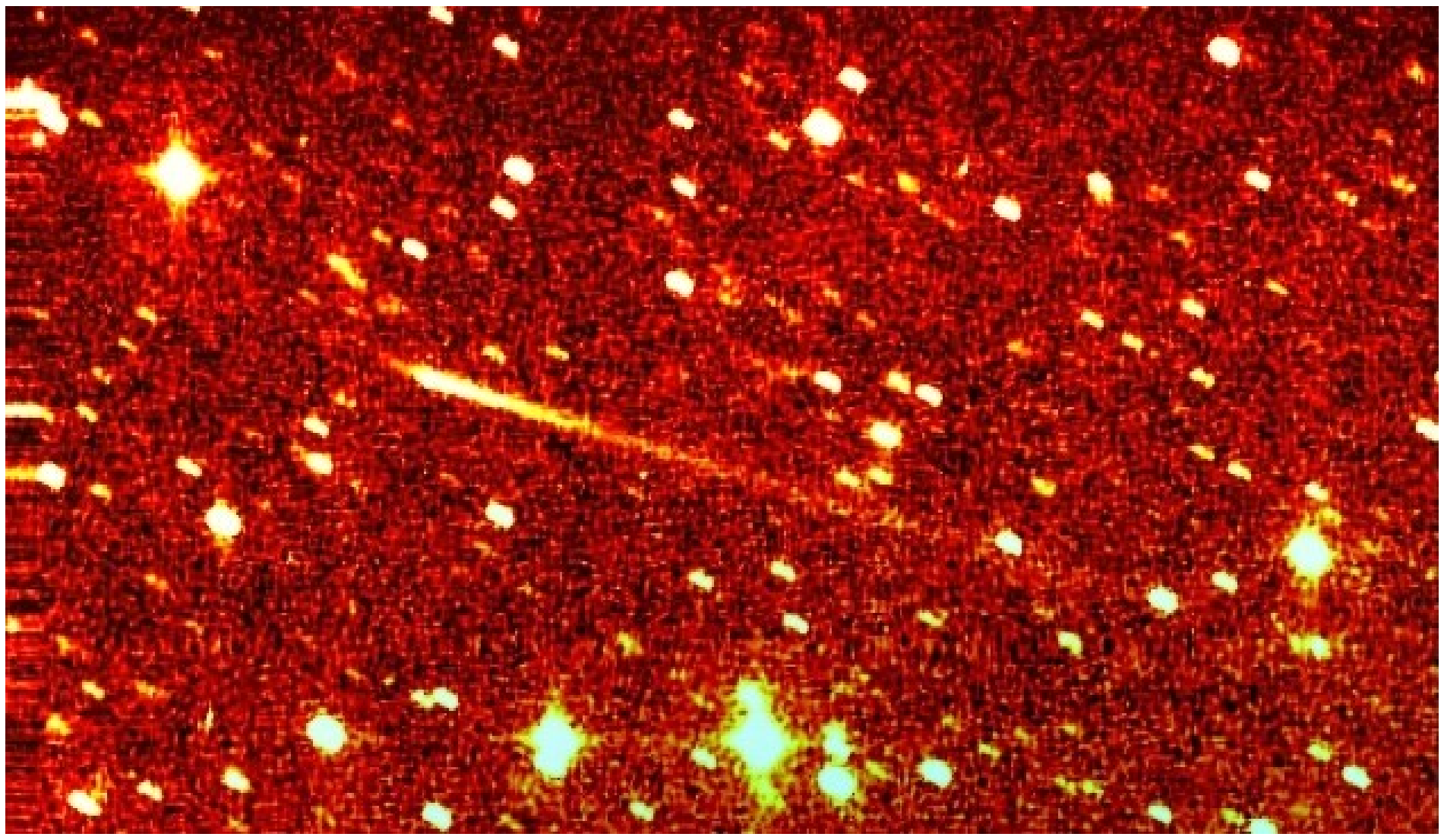

**Figure 2.** Image of 133P/Elst-Pizarro taken by James V. Scotty on September 21st 1996, with the 0.91 m Spacewatch telescope. It exhibits its star-like nucleus with a thin, long, persistent gas-less tail. The small extension toward the upper left is an anti-tail due to the Earth crossing the orbital plane of the object. Reproduced with permission.

<> <> <> <> <> <>

133P/Elst-Pizarro belongs to the Themis family of asteroids (Hsieh et al., 2004) and has been extensively studied (Hsieh, 2004; 2009; Ferrín, 2006; Licandro et al., 2011; Jewitt et al., 2014). We will create the phase plot and the secular light curve of this object.

**2.2 133P/Elst-Pizarro: Phase and Secular Light Curve plots**

In analyzing the data of 133P an important plot is the phase plot (Figure 3). This compares de absolute magnitude $m_V(1,1,\alpha)$ vs phase angle, $\alpha$. These parameters are related by

$$m_V(1,1,0) = m_V(\Delta,R,\alpha) - 5 \log(\Delta.R) - \beta.\alpha \quad (1)$$

where $\Delta$ = object-Earth distance, R= Sun-object distance and $\beta$ = phase coefficient . We are using a linear law in the phase plot because the range of $\alpha$ values is small ($\alpha$<25°) and thus a more sophisticated law is not justified. We use the Minor Planet Center data base to determine the phase law. The phase curve (Figure 3) does not show a dependence with $\alpha$. It looks like the comet is surrounded by an optically thick dust cloud that is masking the nucleus.

The SLC of 133P (Figure 4) shows normal activity with Asec = -2.4±0.1 mag from t-Tq = -80±4 d pre- to t-Tq = 230±4 d post-perihelion for a total active time of 310±6 d . Interestingly after t-Tq ~+230 d and up to ~+550 d the comet exhibits a faint residual activity that we are identifying with CO or CO2 sublimation. Because CO or CO2 reside deep inside the nucleus, it takes some time for the thermal wave to penetrate de surface and reach to the CO or CO2 layer. Thus these two components tend to appear late in the apparition, well past perihelion. This can be clearly seen in Figure 9 of a work by Feaga et al. (2013). They studied comet C/2009 P1 and measured the production rate of H2O and CO. While H2O peaks around perihelion, CO continues increasing at perihelion and well past perihelion. At the last observation 100 days after perihelion and at +2 AU from the Sun, the CO production rate was still growing. Thus there is some basis to believe that the residual activity of 133P might be due to CO or CO2.

**2.3 Comet 133P: Absolute magnitude, diameter, mass**

From the phase and SLC plots (Figures 3 and 4) we estimate an absolute magnitude $m_V(1,1,0) = 15.83\pm0.05$. Next we transform the geometric albedo in the red band $p_R = 0.05\pm0.02$ from Hsieh et al. (2009), to the visual region using our own calibration $p_V = p_R - 0.003 = 0.047\pm0.02$. Jewitt (1991) gives the following formula to determine the diameter:

$$p_V \, .r_{NUC}^2 = 2.24x10^{\wedge}22 \, .R^2 \, .\Delta^2 \, 10^{\wedge}[0.4(m_{SUN} - m_V(1,1,0))] \quad (2)$$

which can be written in a more amicable form

$$\text{Log} [ \, p_V.D^2/4 \, ] = 5.654 - 0.4 \, m_V(1,1,0) \quad (3)$$

Then the diameter of the nucleus is $D_{NUC} = 4.2\pm0.8$ km and $r_{NUC} = 2113\pm400$ m. Adopting their density $\rho \sim 1.300$ kg/m$^3$ we find a mass $M_{NUC} = 5.1x10^{\wedge}13$ kg. These values agree very well with those given below by Jewitt et al. (2014).

Jewitt et al. (2014) find an absolute magnitude Hv=15.70±0.10 with projected dimensions of 3.6x5.4 km (axis ratio 1.5 to 1). They use the IAU formalism, so the absolute magnitude is 0.22-0.3 brighter that our result due to the contribution of the opposition effect. They find an active period of 60-150 days with a mean dust production rate $0.2 < dM/dt < 1.4$ kg/s. They also find a minimum density of $\rho \sim 1.300$ kg/m$^3$ for the nucleus to remain gravitationally bound. Using a measured albedo $p_R = 0.05\pm0.02$ from Hsieh et al. (2009) they calculate an equivalent radius $r_{NUC} = 2.2\pm0.4$ km. Using the above density they find a mass for the nucleus $M_{NUC} \sim 4x10^{\wedge}13$ kg.

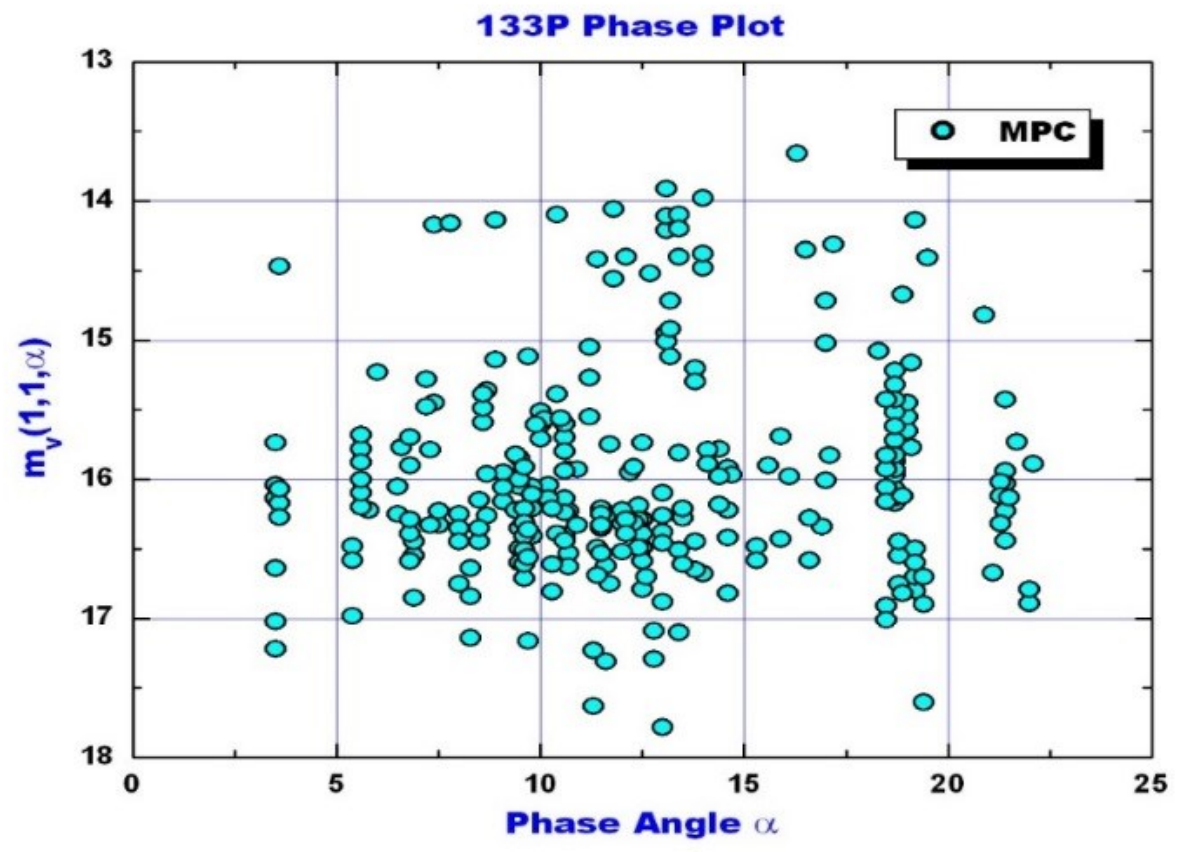

**Figure 3.** The Phase plot of 133P/Elst-Pizarro does not show any phase effect. This can only be understood if the nucleus is covered by a cloud of dust that is optically thick, screening the surface. Thus no phase correction has been applied to our data.

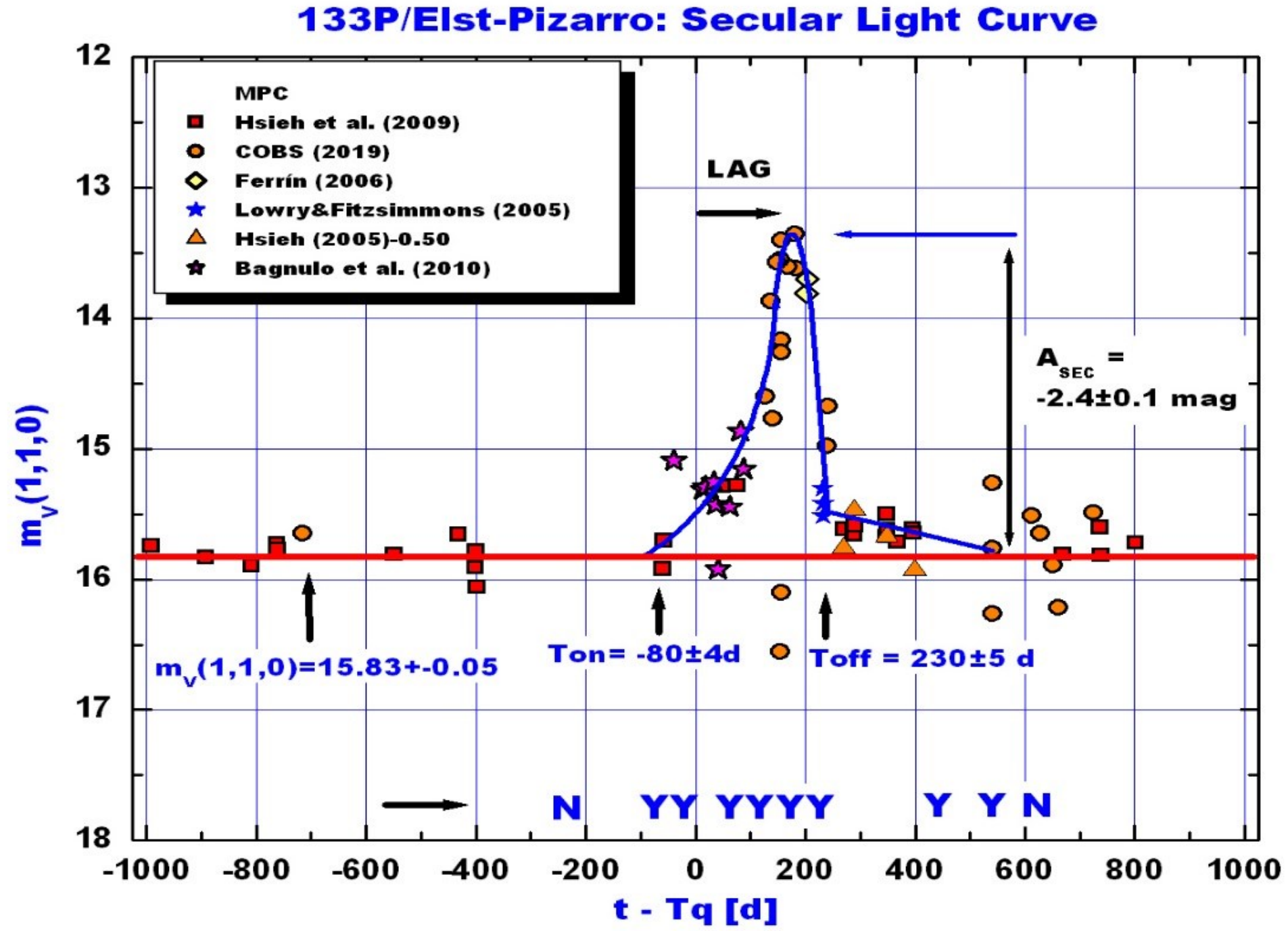

**Figure 4.** The SLC of comet 133P/Elst-Pizarro is quite well defined and it didn´t need help from the MPC database to be determined. The absolute magnitude is very well defined at a level of $m_V(1,1,\alpha) = 15.83\pm0.05$. The LAG time is 170±10 d supporting the idea that this comet has its pole tilted. Prialnik and Rosenberg (2009) were able to reproduce this feature when considering large tilt angles in their simulations. Notice the long tail from day ~+230 to day ~+550 which we conjecture as due to CO or CO2 sublimation. Because CO and CO2 are abundant at deeper levels, it takes more time for the thermal wave to reach there and sublimate. Thus the CO and CO2 contribution tends to appear late into the apparition, after perihelion. Notice the good agreement between the light curve and tail sightings (Y).

<> <> <> <> <>

### 3.1 249P/LINEAR: Introduction

249P/LINEAR is a Jupiter family comet with perihelion distance q = 0.497 AU, aphelion distance Q = 5.601 AU, eccentricity e = 0.82, demonstrating that the condition of a pinpoint comet is not restricted to the Main Belt. An image of 249P/LINEAR exhibiting its thin, long gas-less tail in 2006 is shown Figure 5. Additional images can be found at the following link: http://www.aerith.net/comet/catalog/0133P/2013-pictures.html

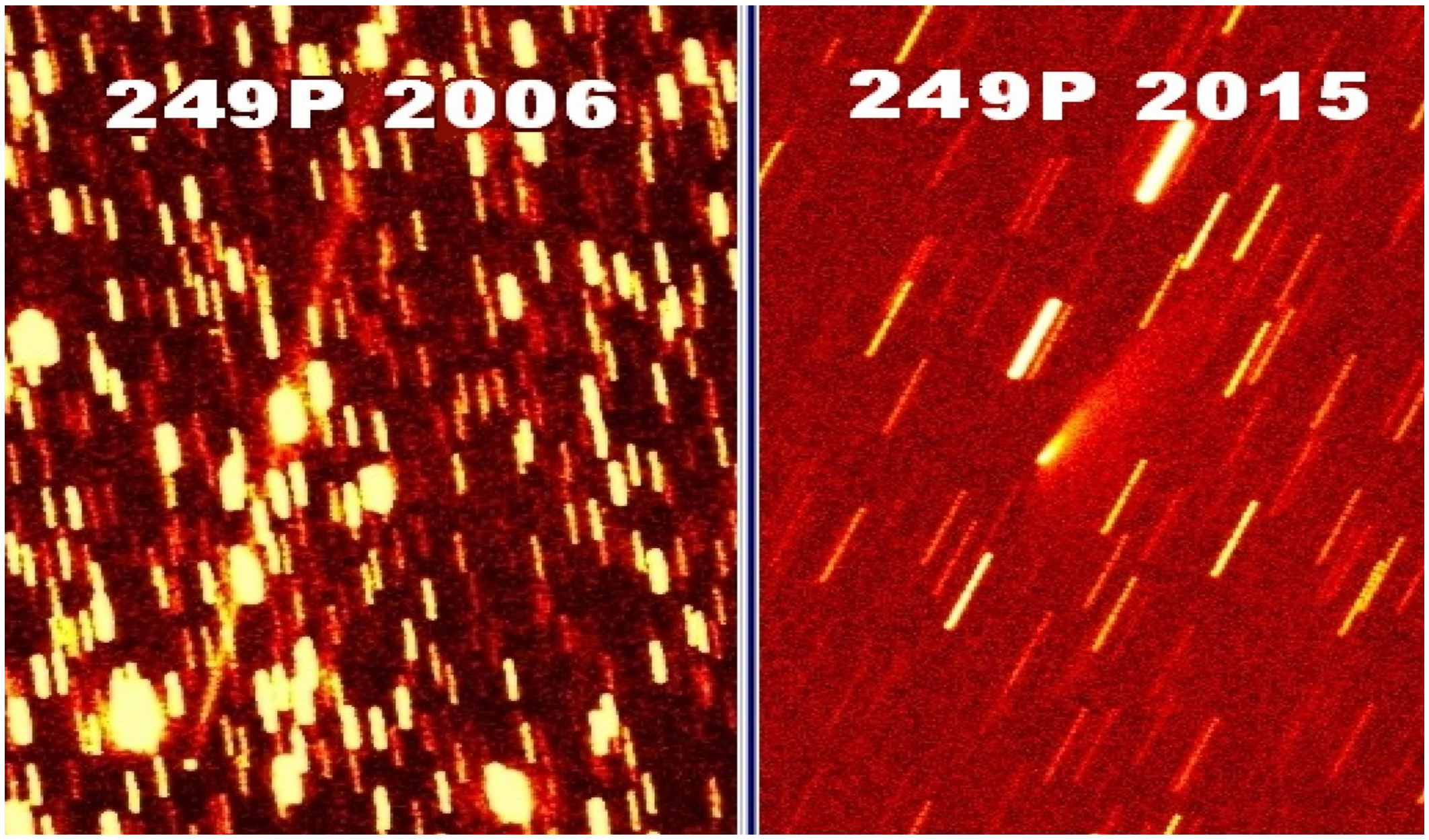

**Figure 5.** Images of 249P/LINEAR in 2006 and 2015. The perihelion distance decreased in this interval from q(2006)= 0.5106 AU to q(2015) = 0.4985 AU, representing an increase in solar insolation by 4.9%. By 2020 q will reach to q=0.4967 AU, for a total increase of 5.3% in received solar energy. This small value was sufficient to move 249P from the PCG into the group of bona fide comets, as depicted on the image on the right hand side. In 2015 it exhibited a non-stellar nucleus with coma, and a tail that expands with distance, clear indication of normal cometary activity. Images: (a) Michael Jager and Gerald Rhemann, November 19th, 2006 with a 20 cm Astrograph, 8 images x 170 s = 22.7 min of total exposure time. (b) Taken by Michael Jager on December 26, 2015, 25 cm telescope, 7 images x 90 sec = 10.5 min. (Reproduced with permission). Perihelions were on Tq = 2006 08 28 and Tq = 2015 11 27. Notice that the tail is detached from the nucleus, implying a turn off of an active region. Thus 249P becomes the best example of a Lazarus comet, one that has become rejuvenated after a diminution of its perihelion distance resulting in an increase in solar insolation, actually a rejuvenation (Ferrín, 2014).

<> <> <> <> <> <>

### 3.2 249P/LINEAR: Phase and Secular Light Curve plots

As mentioned in Section 2.2 the phase plot in combination with the SLC is the best way to determine the absolute magnitude of a comet or an asteroid. In this regard the phase plot of 249P is rather surprising because in 2006 it shows a phase effect but in 2015 the phase effect is gone (Figure 6).

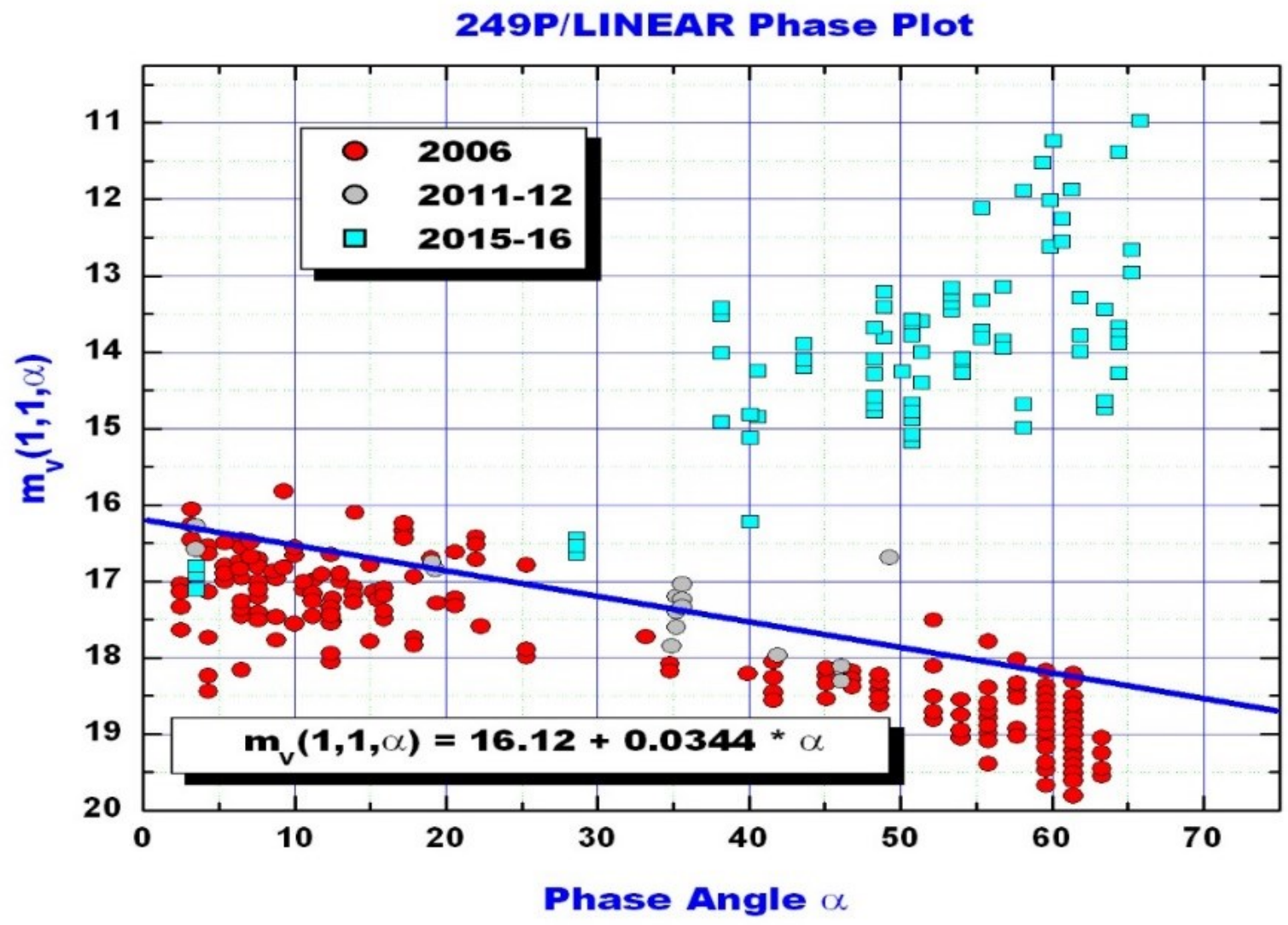

**Figure 6.** The phase plot of 249P/LINEAR exhibits a conflicting result per apparition. A phase plot with a random distribution means that we are not seeing the nucleus and that it is covered with a cloud of dust masking the surface. This is what is happening on 2015-16. On 2006 and on 2011-12 the object was a pinpoint comet and the nucleus was detected with a phase effect. In 2015-16 the comet left the PCG and became a bona fide comet. From the above plot we can read an absolute magnitude $m_V(1,1,0) = 16.13\pm0.05$ and a phase coefficient $\beta = 0.0344\pm0.0020$.

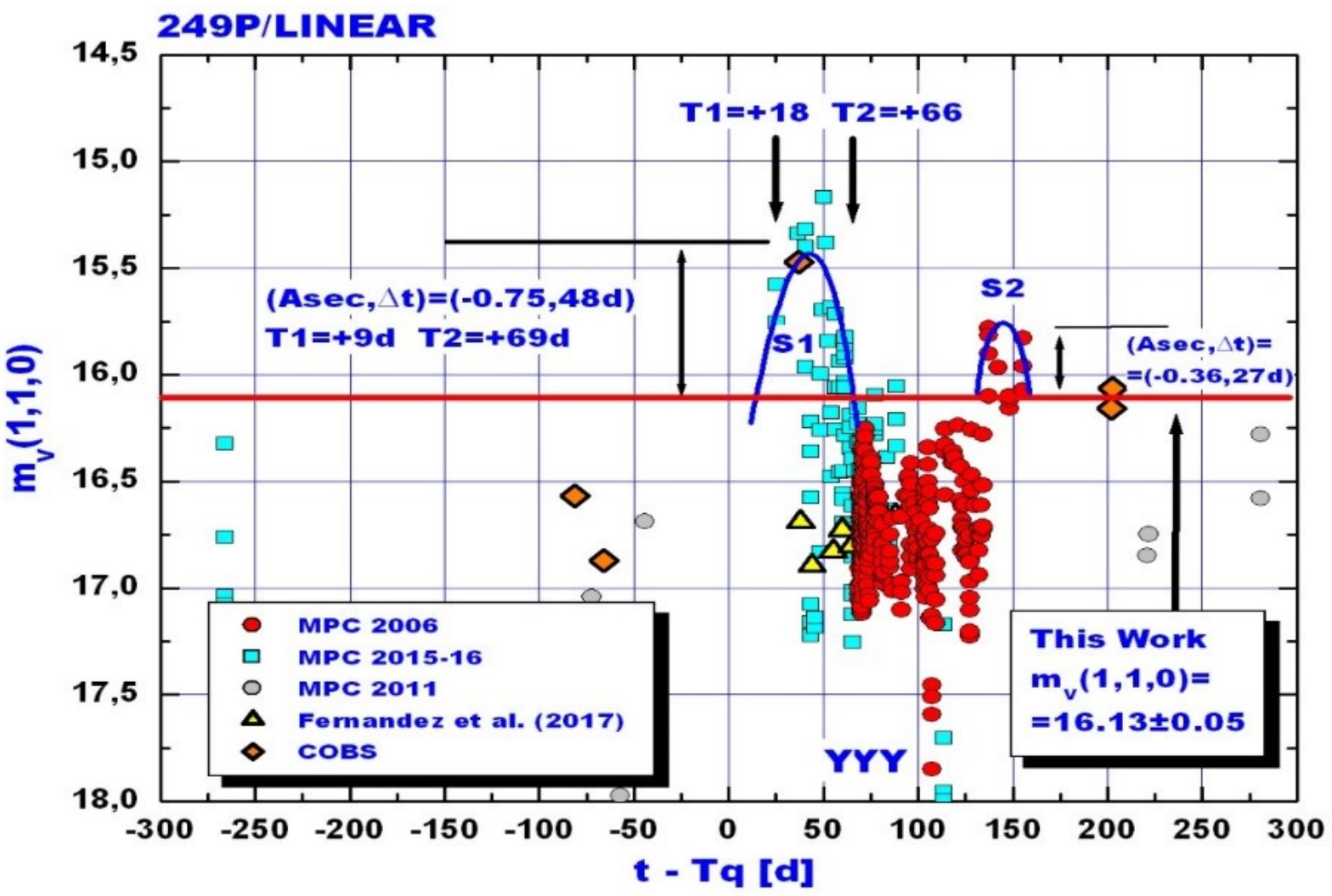

**Figure 7.** The SLC of 249P/LINEAR exhibits two periods of activity, S1 and S2, and evidence of what might be an eclipse. The tail sightings (Ys below) coincide with the period of activity. The selection of the absolute magnitude was in part dictated by the two COBS data points at t-Tq = +200 d whose absolute magnitude coincides with the one derived from the phase plot. The eclipse region is enlarged in the next Figure. There should be another eclipse half the orbital period away, but the region ~ -275 d is devoid of data.

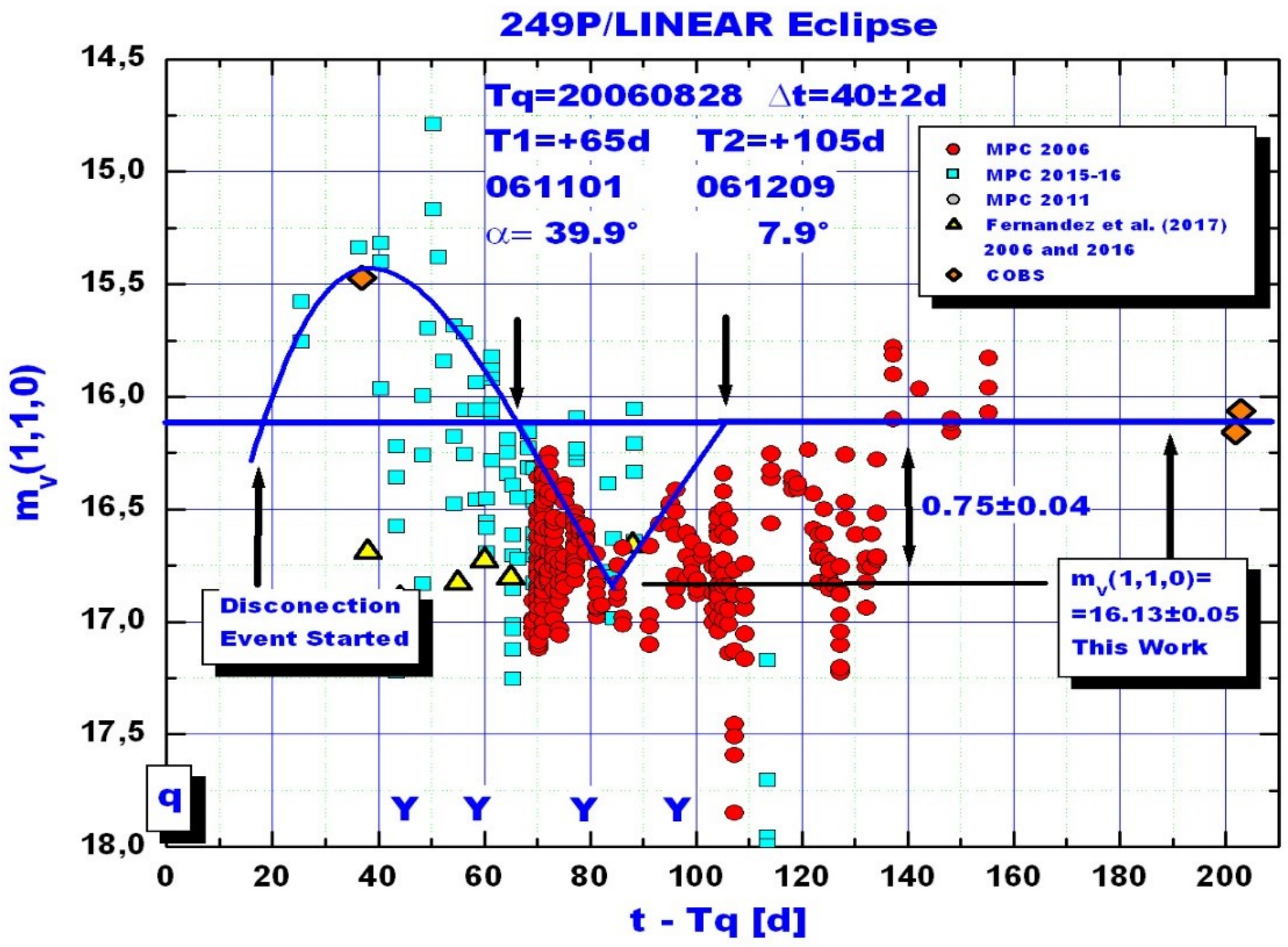

**Figure 8.** Region of the eclipse enlarged. When two same size spherical components eclipse each other the depth of the eclipse is 0.75 magnitudes, and this is the maximum value the eclipse can have. Since this is what is observed in this plot, we conclude that 249P is composed of two identical members. Other parameters of the eclipse are derived in the text. On 2015-2016 the eclipse was masked by the cometary activity.

<> <> <> <> <>

The comets with decreasing perihelion distance have been designed as Lazarus comets (Ferrín et al., 2013). A good example of a Lazarus comets is 107P/Wilson-Harrington and now 249P/LINEAR.

### 3.3 Comet 249P: Absolute magnitude, diameter, mass

There is only one paper in the literature dealing with this comet. Fernandez et al. (2017) find an absolute magnitude Hv=17.0±0.4 with a nuclear radius $r_{NUC}$ = 1.15±0.35 km and an active period of only ~20 days around perihelion. They find a similar pattern of activity in 2006 and 2016. According to them, the activity onset, duration and total ejected mass were very similar during the 2006 and 2015 perihelion passage. They found a maximum dust ejection rate taking place 1.6 d before perihelion. They calculate a total ejected mass (2.5±0.9)x10^8 kg in the previous passage.

Our results differ significantly from the above values. The phase plot and the SLC (Figures 6 and 7) show very different photometric behaviors in 2006 and 2015-16, including different mass loss values. From the SLC our absolute magnitude is $m_V(1,1,0)$ = 16.13±0.05. We find two sources of activity, S1 starting at t-Tq = +20 days and S2 starting at t-Tq = +131 days, with evidence of an eclipse in the middle.

Since this is a Jupiter Family comet it is reasonable to adopt the geometric albedo suggested by Kokotanekova et al. (2017), pv = 0.042±0.004. Using Equation (2) we then find a diameter $D_{NUC}$ = 3.89±0.32 km and a radius $r_{NUC}$ = 1950±240 m. Adopting a minimum density suggested by Hsieh et al. (2004) , $\rho_{min}$ = 1300 kg/m$^3$, we get a mass $M_{NUC}$ = (4.0±0.6)x10^13 kg. The calculation of the diameter assumes that the nucleus is a single object. However in Figure 8 we learn that the comet might be double.

**3.4 249P/LINEAR: Eclipse**

Figure 8 shows the eclipse feature of 249P and in Figure 9 we appreciate the eclipse geometry of a pair of identical size nuclei. The maximum depth of an eclipse of two identical objects is 0.75 magnitudes. Since this is what is shown in Figure 8, we have to conclude that 249P is double and composed of two identical members. We are going to calculate the parameters of the system.

The absolute magnitude of 249P from this work is $m_V(1,1,0)$ = 16.13±0.05, or $m_V(2)$ = 16.13. This is the combined light from the two objects $2\pi r^2$. The light from one single component $\pi r^2$ is 0.75 magnitudes fainter, $m_V(1)$ = 16.88. Assuming geometric albedos of 0.042 (Kokotanekova, 2017), 0.06 and 0.075 (Hsieh et al., 2009), we find diameters D = 2.76 km, 2.31 km, 2.06 km, and radius r = 1378 m, 1153 m, 1031 m, and assuming a density of 600 kg/m$^3$ (Kokotanekova, 2017), we find masses 6.57, 3.9, 2.7x10^12 kg.

We are going to interpret this eclipse as a solar eclipse. From the JPL Horizonts tool we calculate the ecliptic longitude L of the beginning and end of eclipse, Linit = 55.0° and Lend = 70.4°. Thus angle γ in Figure 9 measures γ = 7.7°. Assuming a circular orbit of radius a we have

$$r / a = \tan \gamma \quad (4)$$

and using Kepler's third law

$$P_{orbit}^2 = 4 \pi^2 a^3 / [ G (m_1 + m_2)] \quad (5)$$

we calculate a = 10610 m, 8877 m, 7938 m, and Porbit = 2.52 d, 3.30 d and 3.90 d.

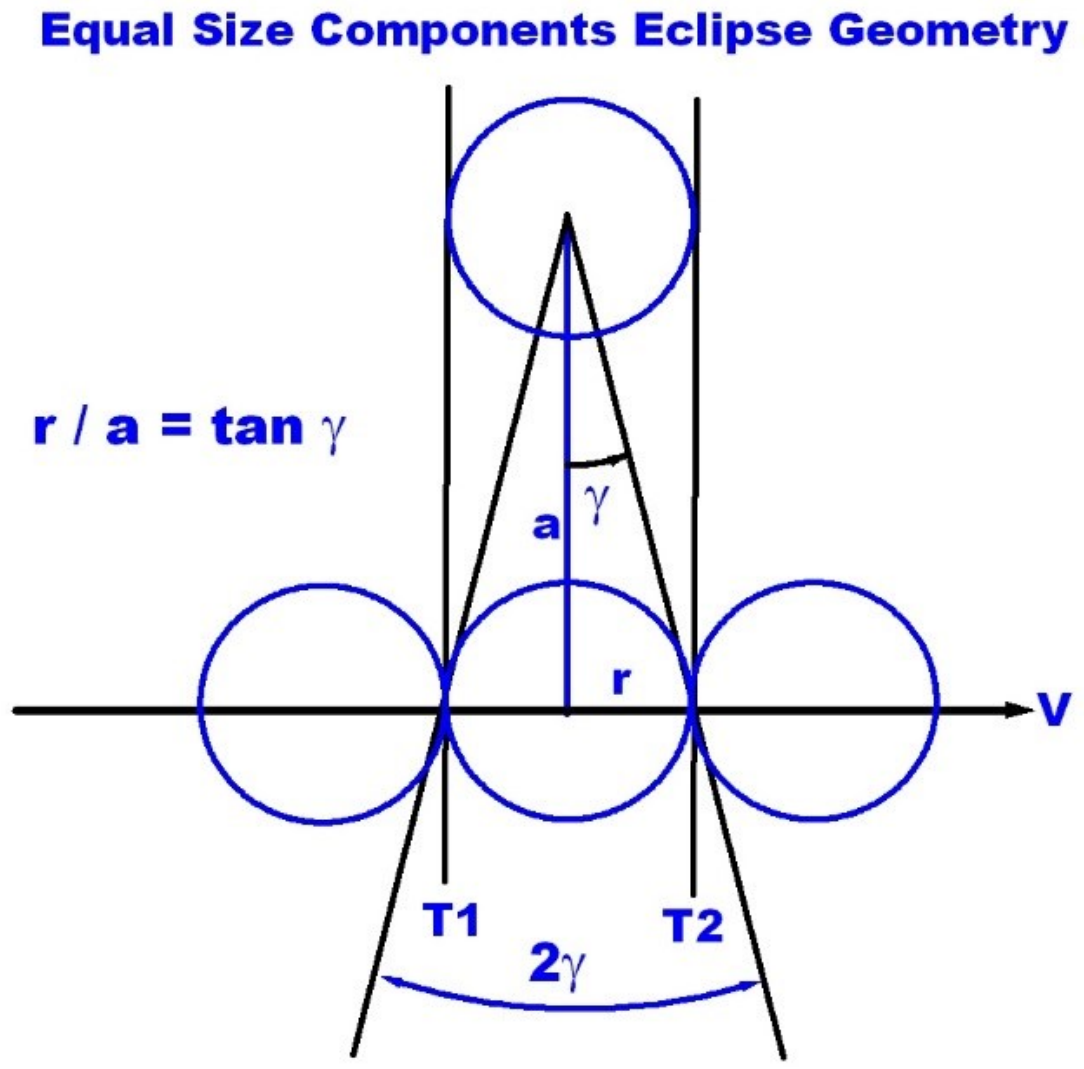

**Figure 9.** Eclipse geometry. With this information it is possible to calculate the parameters of the components.

### 3.5 Solar Energy Budget

In Figure 10 we show the solar insolation of 249P. It is given normalized to 2006 and it is a constant divided by the solar distance squared.

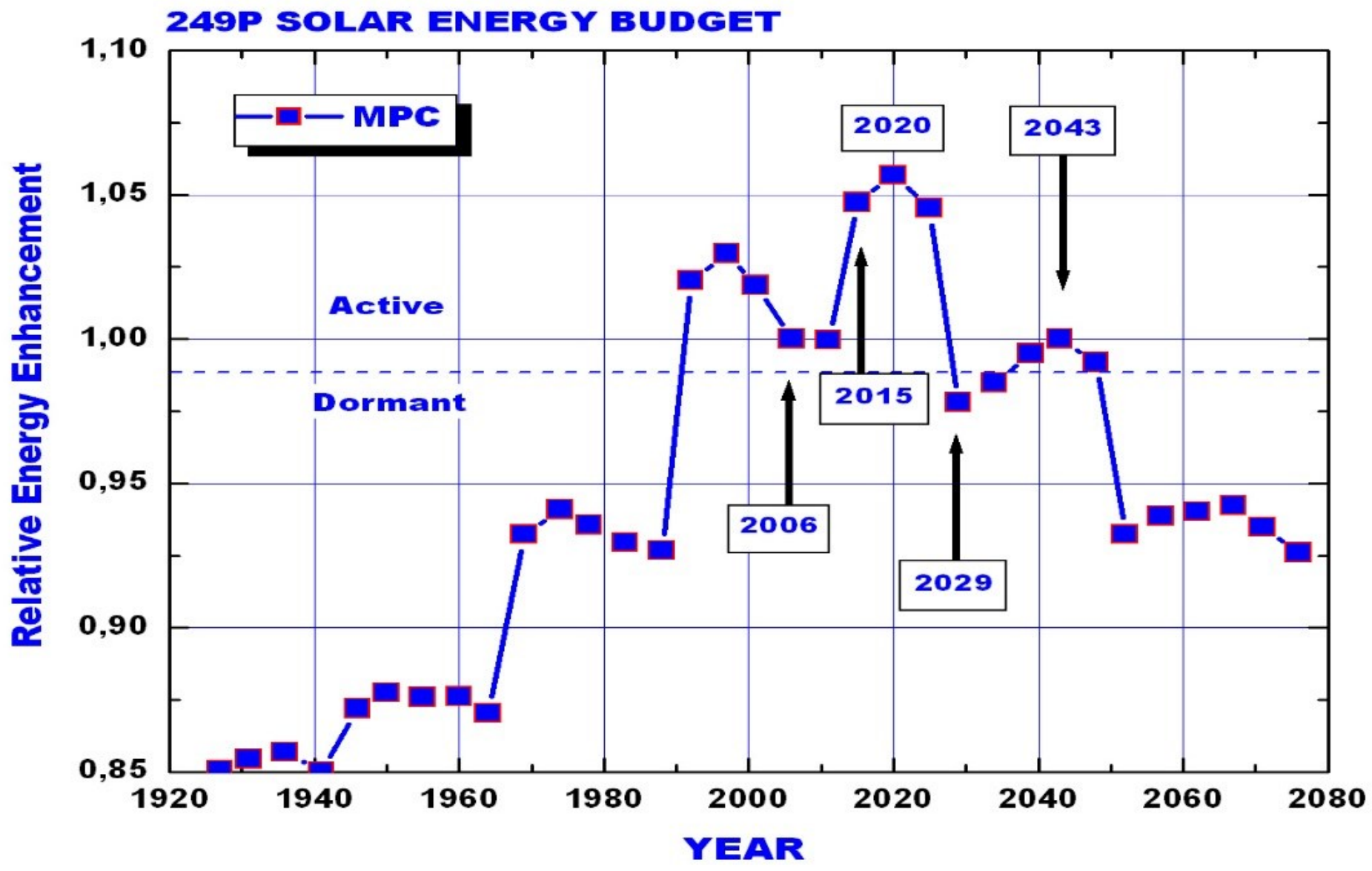

**Figure 10.** Solar insolation calculated for 249P. The values are normalized to the budget of 2006. From 2006 to 2015 the budget increased by 4.9% and that value was sufficient to move out 249P from the PCG phase into the bona fide comet phase. Interestingly this situation is not going to last for long. On the year 2029 the insolation will return to values below that of 2006 when the activity existed but was feeble. We will be able to see how the comet responds to this diminution.

### 4.1 331P/Gibbs: Introduction

There is little information about this comet in the literature. The NASA Ads search query does not list any manuscript with that name. The Minor Planet Center gives a perihelion distance of q = 2.88 AU and an eccentricity value e= 0.041, so the orbit is practically circular with the aphelion distance well inside the belt, Q=3.12 AU. An image of 331P/Gibbs is shown in Figure 11. Additional images can be found here http://www.aerith.net/comet/catalog/0331P/pictures.html

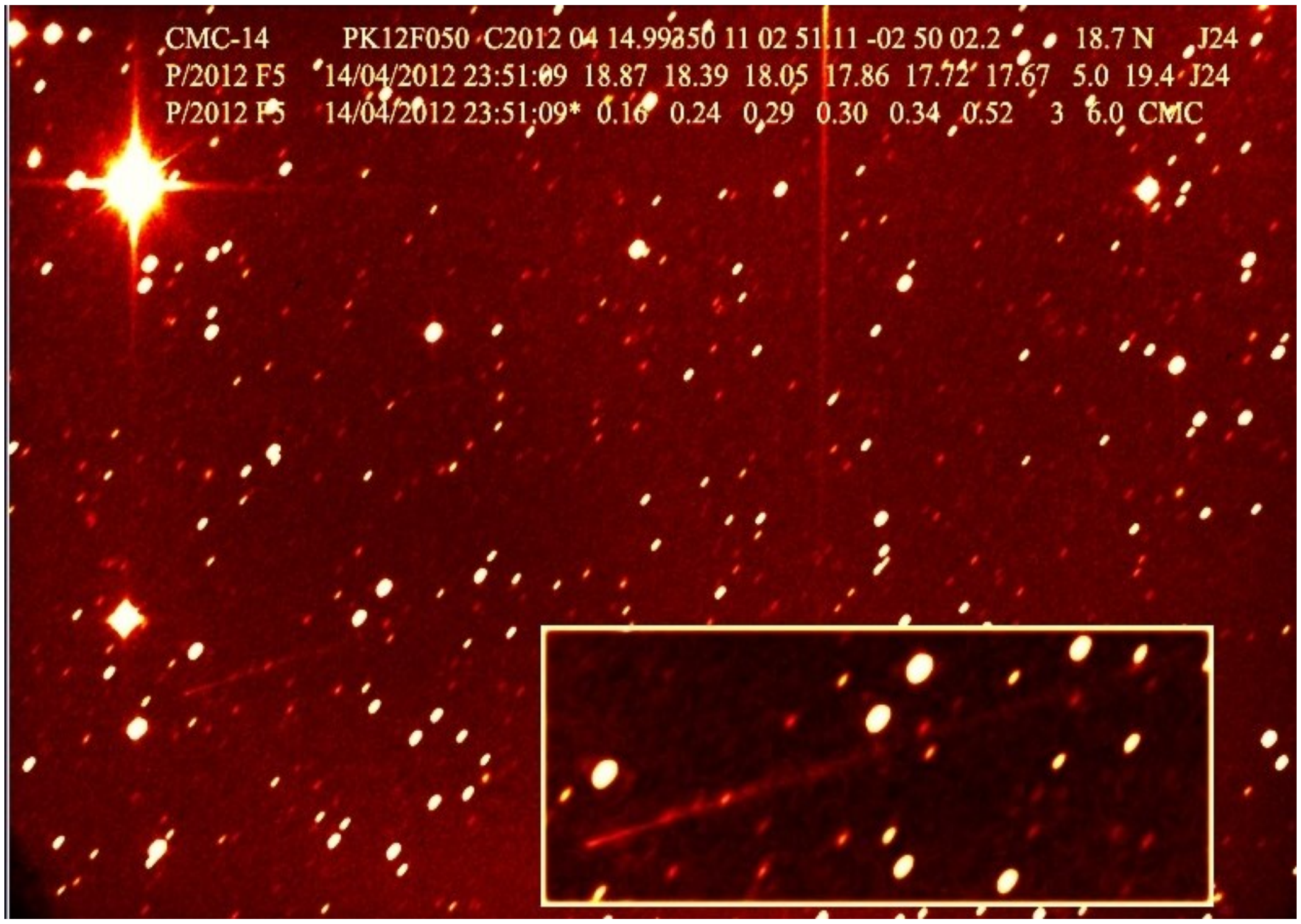

**Figure 11.** Image of 331P/Gibbs taken by Jose Francisco Hernandez with a 40 cm telescope on the 14$^{th}$ of April, 2014. Reproduced with permission.

<> <> <> <> <>

### 4.2 Comet 331P/Gibbs: the phase and SLC plots

The phase and SLC plots are shown in Figures 12 and 13, and an enlargement of the SLC is shown in Figure 14. This comet has very few observations, 70 in total, so the phase plot and the SLC are poorly defined. The SLC shows a region of activity at +700 to +750 days after perihelion with an amplitude Asec = -0.83±0.05 magnitudes. The activity lasted for only 60 days. The obvious recommendation is to get additional observations all along the orbit to better define the interval of activity. Also the fact that activity takes place so far in time from perihelion may be an indication that this comet has the nucleus polar region tilted toward the Sun at this orbital position like 133P did. However since the orbit is practically circular, this does not imply a significantly larger distance from the Sun.

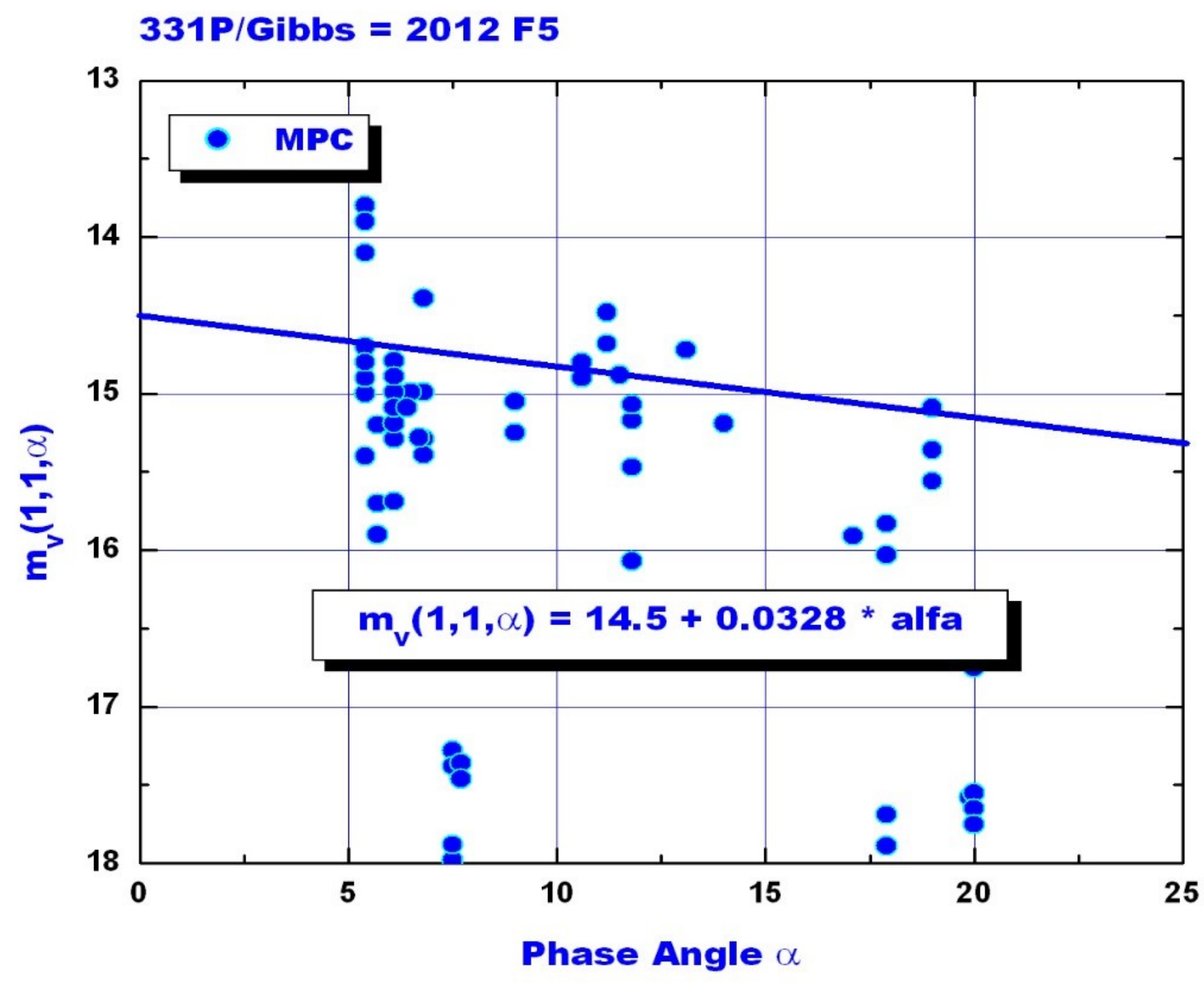

**Figure 12.** Phase plot of 331/Gibbs. Observations with $\alpha < 5.5°$ have been eliminated because the opposition effect produces false positives.

<> <> <> <> <>

## 4.3 Comet 331P/Gibbs: Absolute magnitude, diameter and mass

The best absolute magnitude is reached comparing the phase plot with the SLC plot. Both give information in different phase spaces and have some weight on the final solution. Since the maximum angle in the phase plot is 20° and besides there are few observations only a linear fit can be attempted. We find a slope $\beta = 0.0328 \pm 0.0010$ and an absolute magnitude $m_V(1,1,\alpha) = 14.5 \pm 0.05$.

Using the above magnitude and a geometric albedo $p_V = 0.042$ representative of Jupiter Family comets (Kokotanekova, 2017), we obtain from Equation (1) a diameter D = 8.25 km and a radius r = 4130 m. If we adopt as a density $\rho = 1300$ kg/m$^3$ (the same as 133P) the mass is M =3.8 x 10^14 kg.

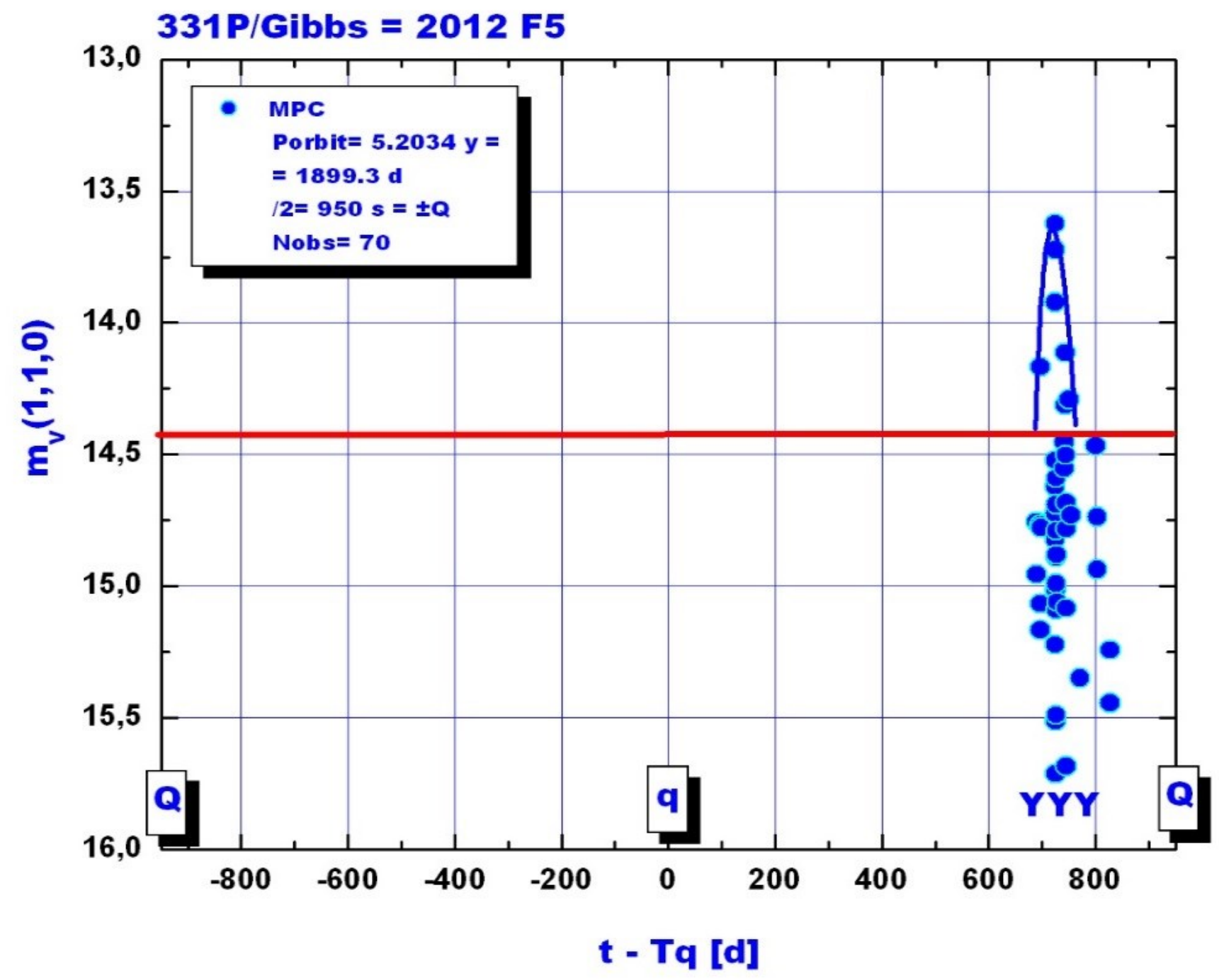

**Figure 13.** The SLC of 331P/Gibbs.

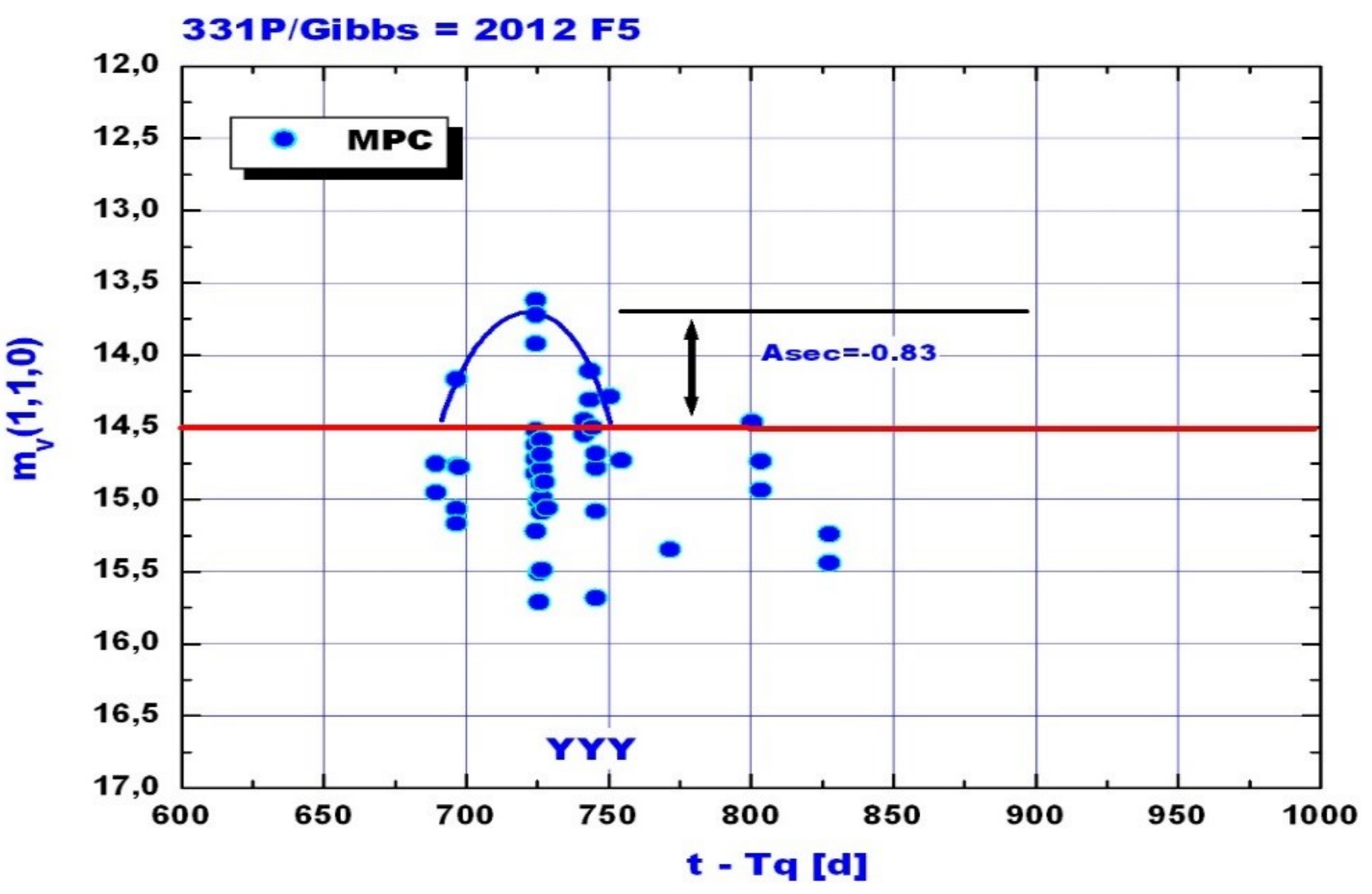

**Figure 14.** An enlargement of the region of activity of 331P/Gibbs. The active region coincides with the region where a tail was observed.

### 5.1 62412 = 2000 SY178: Introduction

62412 is a main belt comet with perihelion distance 2.90 AU and aphelion distance 3.41 AU. The orbital period is 5.59 y and the absolute magnitude listed by the MPC is H = 13.8, similar to the value obtained in this work, $m_V(1,1,0)$ = 14.00±0.05. An image of 62412 in stellar condition is shown in Figure 15. The only image available with a tail is shown in Sheppard and Trujillo (2014).

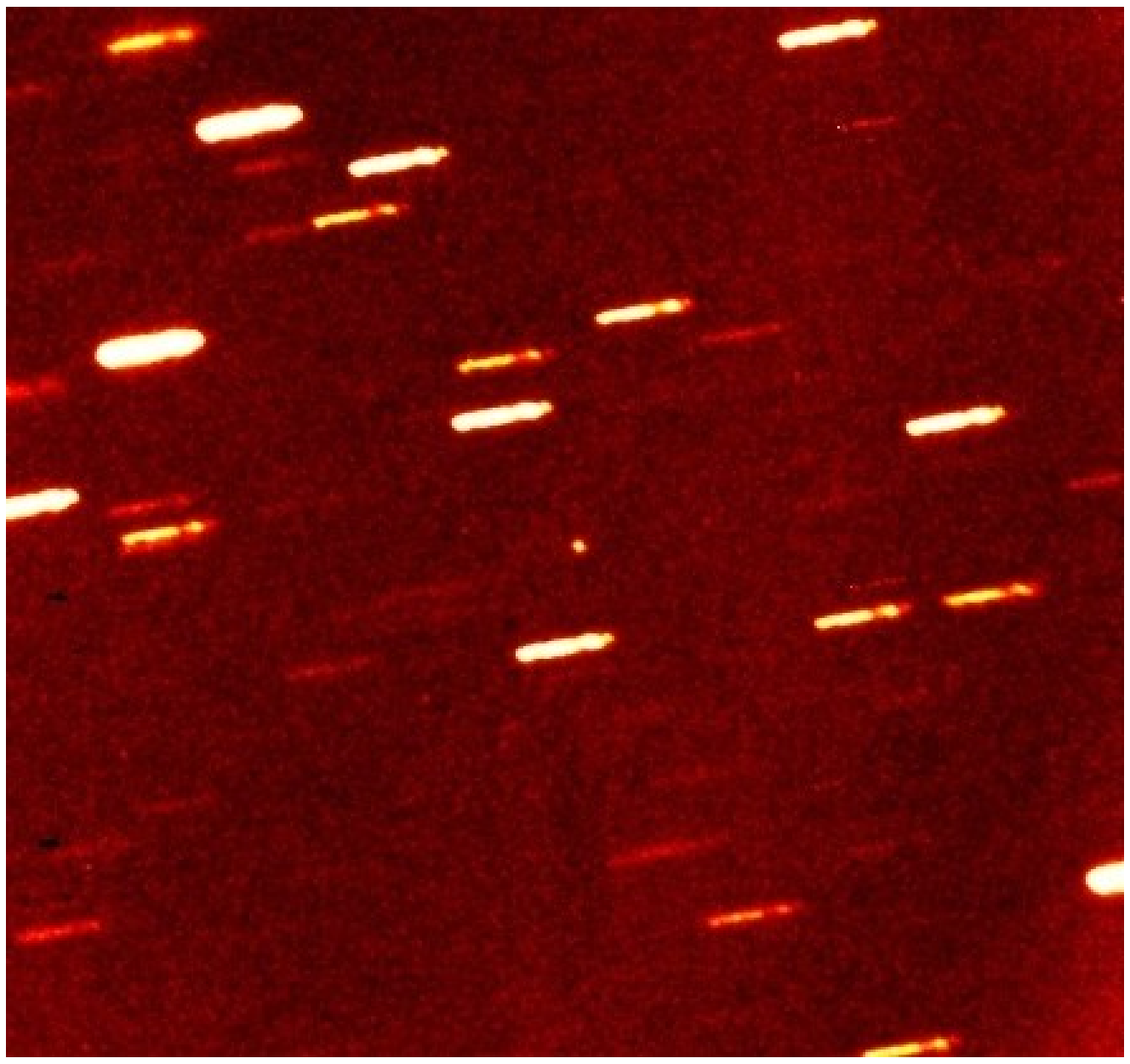

**Figure 15.** Active asteroid 62412 imaged by Agustin Acosta on March 29th, 2019. 24 images of 5 minutes each were averaged and stacked for a total exposure time of 2.0 hours. The AA looks stellar with no coma or tail discerned at this location in the orbit.

<> <> <> <> <> <>

### 5.2 Comet 62412: the phase and SLC plots

The phase and SLC plots are shown in Figures 16 and 17. The phase plot shows a phase effect with slope β = 0.0476±0.0005. The SLC does not show any region of activity in the observed interval, however the region beyond t – Tq > 400 d is poorly covered with data.

### 5.3 Comet 62412: Absolute magnitude, diameter and mass

From the phase and SLC plot we derive an absolute magnitude mv(1,1,0) = 13.90±0.05 using the geometric albedo pv = 0.042. Using Equation (2) we find a diameter D = 10.7±0.6 km and a radius r = 5390 m. Adopting a density of 1000 kg/m3 we find a mass M = 6.5x10^14 kg.

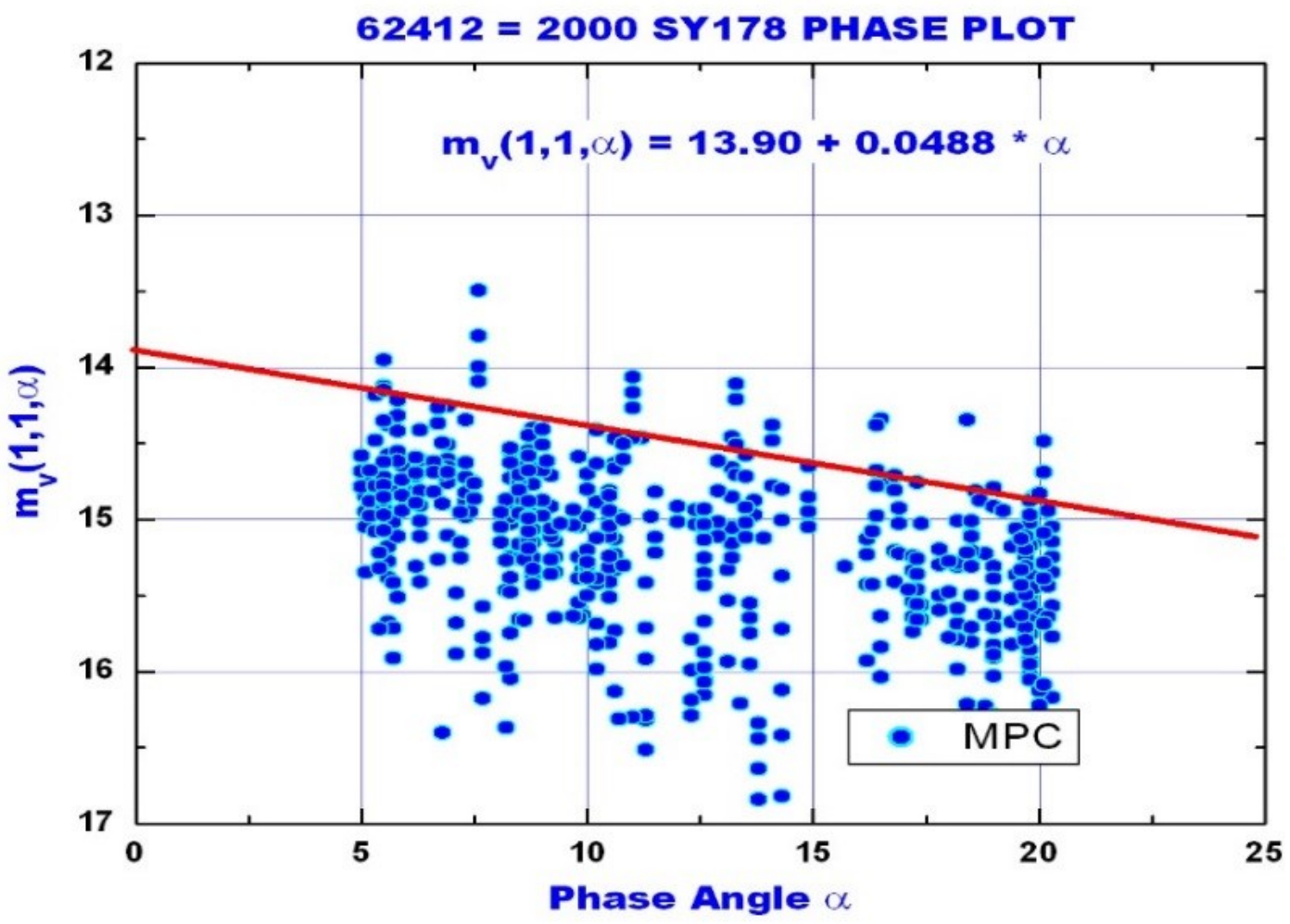

**Figure 16.** Phase plot of 62412. Observations with $\alpha < 5.5°$ have been eliminated because the opposition effect produces false positives.

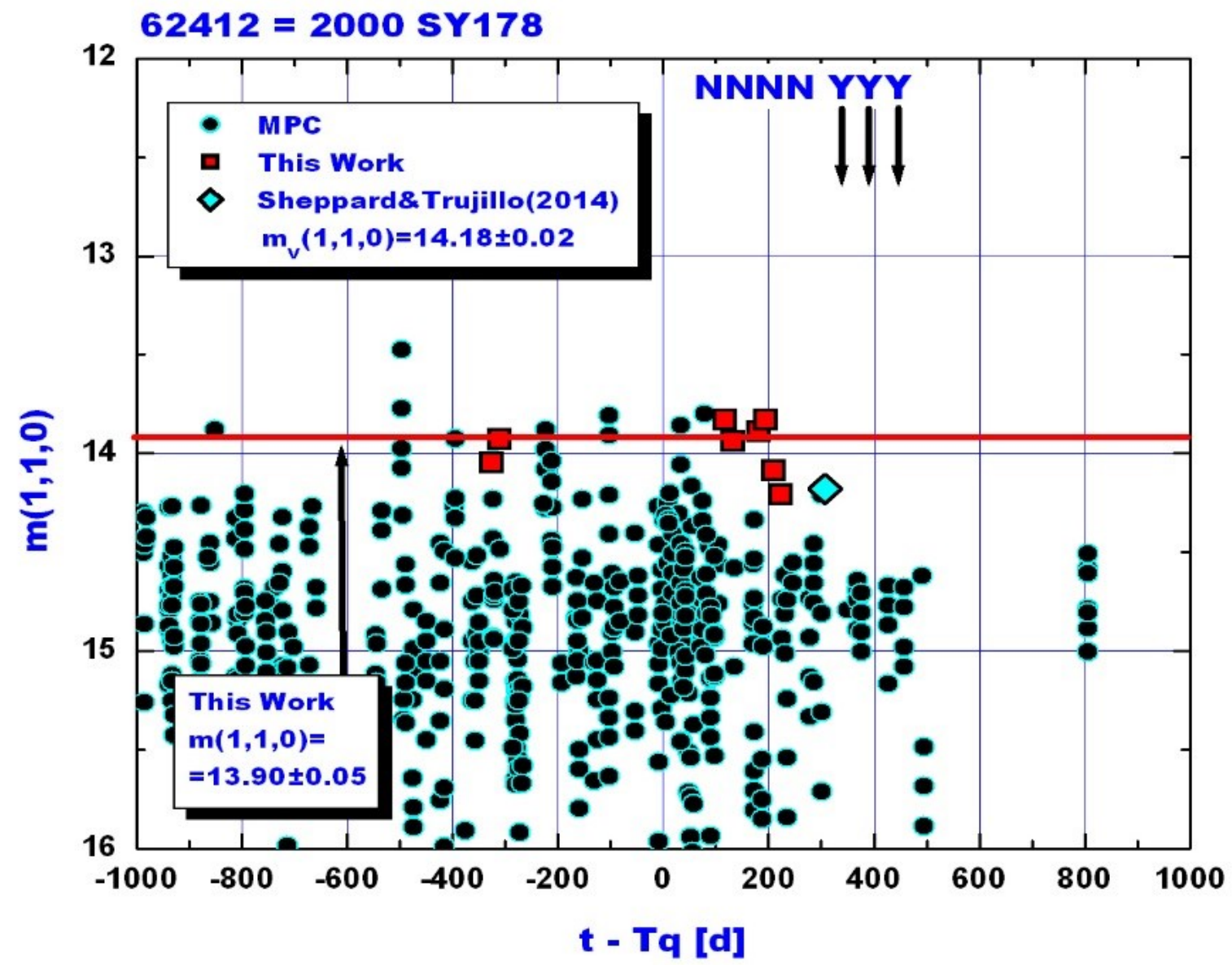

**Figure 17.** The SLC of 62412. The letters above indicate when the tail was (Y) or was not (N) observed on 2014. Neither our photometry nor the MPC photometry exhibit an enhancement on this plot. Thus we have not been able to detect the zone of activity of this comet. Although our observing campaign of 2019 extended to -183 d before the observation of activity detected by Sheppard and Trujillo (2014), the comet did not exhibit evidence of activity or tail. The region t -Tq > +200 d is poorly covered by observations. Our guess is that the activity of this object might be short and feeble. Notice the location of the Sheppard and Trujillo´s data point and the last observation of our photometry less than 180 days away.

**6.1 6478 Gault : Introduction**

6478 Gault is a km-sized asteroid in the Phocaea family (Nesvorny, 2015). The MPC lists the perihelion distance as q=1.86 AU, the aphelion distance 2.75 AU and the eccentricity as e=0.19, thus the orbit is entirely asteroidal. An image of 6478 Gault exhibiting its thin, long gas-less tail is shown in Figure 15. Additional images can be found at the following link: http://www.aerith.net/comet/catalog/A06478/2020-pictures.html.

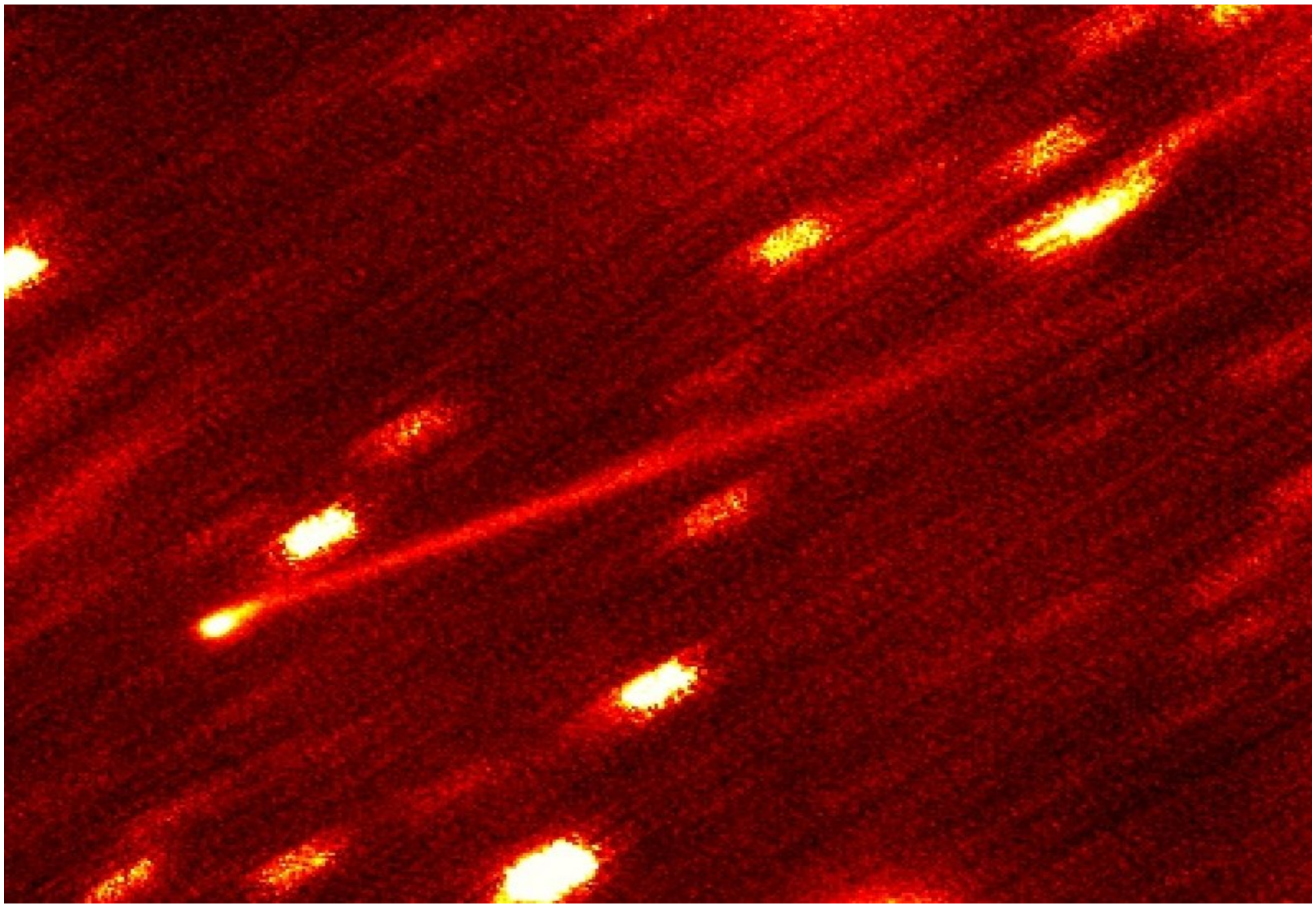

**Figure 18.** 6478 Gault. Median image of 15 frames x 5 min = 75 min of exposure, taken with the 1 m f/5 reflector telescope of the National Observatory of Venezuela, on February 5th, 2019, 00h 48m UT (IF, this work). The characteristic features, star-like head, thin, long, gas-less persistent dust tail, and no tail spread as a function of distance, are all exhibited in this object making it a bona fide member of the group. A second tail shorter and above the main tail is compromised by a star. The rotation period of 3.360±0.005 h determined in this work, confirms membership into the rotational disruption regime.

<> <> <> <>

**6.2 6478 Gault: Resume of our previous work**

6478 Gault has been studied by our group previously so much of the information obtained will not be repeated here, only the main results (Ferrín and Acosta 2019a; Ferrín el al., 2019b). The phase plot of 6478 Gault does not exhibit a phase effect, implying that we are not seeing a surface. The object must be covered with a cloud of dust that hides the nucleus. That is the reason why we have not applied any phase correction to the observational data. The SLC of 6478 Gault is rather surprising. The object shows 6 peaks of activity the reason for which is not clear at the present time. The total mass released in these 6 zones of activity was calculated. The absolute magnitude listed by the Minor Planet Center and the JPL Small Body Database Browser, H= 14.4, is too bright. A new value $m_V(1,1,\alpha) = 16.12 \pm 0.05$ is derived in this work and it is the mean value of our six faintest measurements. Because previous works (Chandler et al., 2019; Hui et al., 2019; Jewitt et al., 2019; Kleyna et al., 2019; Moreno et al., 2019; Ye et al., 2019) used a much brighter absolute magnitude, their model results might have to be revised.

It is possible to calculate the diameter if the geometric albedo is assumed. Kokotanekova et al. (2017) calculated a mean geometric albedo of the Jupiter Family of comets $\langle p_V \rangle = 0.042 \pm 0.005$. Since the SLC of 6478 indicates that the object is behaving more like a comet than an asteroid, we are going to adopt this value in our calculations. Then we find for the diameter D = 3.91±0.25 km, r = 1955±125 m. To calculate the mass we need a density. We found a rotational period for Gault Prot = 3.360±0.005 h and an amplitude Amp= 0.16±0.02, which translates to a ratio of axis a/b = 1.16. The minimum density to hold material at the tip of the rotating asteroid is $P_{CRITICAL} \sim 1120$ kg/m$^3$ (equation 6). Then the estimated mass of the nucleus is $M_{NUC} = 3.3$ x 10^13 kg. This solution implies a single nucleus for the object. However, as we will see, the object exhibits evidence of duplicity and when this is taken into account we find a total mass for the binary M(p+s) = 2.6x10^13 kg (Section 7.7).

Kleyna et al. (2019) and Moreno et al. (2019) had long series of observations to derive a rotational period, but the photometric time series did not exhibit a light curve. Our own attempt was positive. The periodogram of 6478 Gault showed one peak with a periods ~0.07 days corresponding to Prot(1)=1.680±0.005 hours. A second peak with period ~0.23 days was of poor quality and was rejected. This light curve with one rotational peak violates the rotational limit for disruption (~2.2 h), so we adopted a period with two rotational peaks corresponding to Prot(1)=3.360±0.005 hours, similar to the rotational periods of other members of the PCG (Table 4). The amplitude was Amp = 0.16±0.02 magnitudes.

The minimum density needed to ensure the material at the tip of a prolate body in rotation about its minor axis that is gravitationally bound, is approximately given by Jewitt et al. (2014)

$$\rho_{CRITICAL} \sim 1000\,(3.3/Prot)^2\ [a/b] \qquad (6)$$

If we use the observed period, 3.360 h and the observed ratio a/b=1.16 then the minimum density comes out to be $\rho_{NUC} \sim 1.12$ gm/cm$^3$.

Surprisingly the rotational light curve exhibited a deep in the shape of a flat bottom V that is interpreted as an eclipse. To produce a flat bottom the two components have to have different sizes with radii of the primary and secondary, $r_P$ and $r_S$. Assuming spheres and equal albedos ($p_V$ = 0.042, 0.06, 0.075) we find a radius of the primary $r_P$ = 1850 m, 1550 m, 1390 m, and for the secondary $r_S$ = 660 m, 552 m and 495 m. In order to get these results a minimum density of $\rho_{MIN}$ = 1590 kg/m$^3$ has to be adopted. If the density increases above this value the pair is detached (Ferrín and Acosta 2019a; Ferrín el al., 2019b).

## 7. Evolutionary diagram
### 7.1 Introduction

We present an evolutionary diagrams, the Remaining Returns vs Mass Loss-Age, RR vs ML-Age, introduced to attempt to compare comets in different stages of evolution (Ferrín, 2012). The diagram is shown in Figure 20. It is interesting to question where on this diagram the PCG lies. Do they concentrate on a region or are they all scattered around the plot? However, before we answer this questions it is important to acquaint ourselves with the meaning and interpretation of this complex diagram.

### 7.2 The Remaining Returns vs Mass-Loss-Age diagram

If r = radius of the nucleus, then on the *vertical axis* we shows RR = radius of the nucleus / Δ radius lost per return:

$$RR = r/\Delta r \qquad (6)$$

It is possible to calculate this quantity from the total mass of the nucleus divided by the mass lost per apparition:

$$M/\Delta M = 3 \quad \text{thus} \quad r/\Delta r = 3\ M/\Delta M \qquad (7)$$

On the *horizontal axis* we plot ML-Age given as function of the Mass-Loss Budget. It is reasonable to choose ML-Age as a proxy for age for comets because as they age their mass output decreases with time.

$$\text{ML-Age [cy]} = 3.58\text{x}10\text{^}11\ \text{kg / ML-Budget} \qquad (8)$$

where cy stands for comet years that should not be confused with Earth's years. The constant was chosen so that comet 28P/Neujmin 1 has a ML-Age = 100 cy. This constant will be recalibrated in the next apparition of this comet on 2020.

Why was r/Δr chosen as the significant parameter of the evolutionary plot? The reason is that for a purely sublimating comet, the thickness of the layer removed per apparition should remain approximately constant as a function of time, as can be seen from the following argument.

The energy captured from the Sun depends on the cross section of the nucleus, $\pi.r^2$, on the Bond Albedo, $A_B$ and on the solar constant, S. The energy conservation equation can be written

$$(1 - A_B).S.\pi.r_{nuc}^2 = p_{IR}\ .\sigma.T^4 + k1.4.\pi\ r_{nuc}^2.\Delta r_{nuc}.L + k2.\partial T/\partial x \qquad (9)$$

where $r_{nuc}$ is the nucleus radius, $p_{IR}$ the albedo in the infrared, T the temperature, k1 and k2 are constants, σ the Stefan-Boltzman constant, L the latent heat of sublimation, $\Delta r_{nuc}$ the thickness of the layer removed, and x the depth below the surface.

The term on the left is the energy captured from the Sun. The first term on the right is the energy re-radiated to space, the second the energy sublimated, and the third the energy conducted into the nucleus. The first and third term on the right hand side are small compared with the second term because at large distances temperature is low. The second term dominates near the Sun at perihelion. So, as a first approximation we have

$$( 1- A_B ).S.\pi.r_{nuc}^2 \sim k1.4.\pi.r_{nuc}^2.\Delta r_{nuc} .L \quad (10)$$

$$\Delta r_{nuc} \sim (1 – A_B ).S / ( 4.k3.L) \quad (11)$$

Thus we find that $\Delta r_{nuc}$ should be approximately constant for a sublimating-away comet. Notice that then r/Δr would tend to zero as the comet sublimates. However if the comet contained much dust, part of the dust will fall back on the surface and Δr will tend to zero owing to suffocation, and then r/Δr will tend to infinity.

As a consequence the RR vs ML-Age plot is divided into two large regions, half above and half below. For comets that sublimate away Δr is about constant and r decreases as a function of time. Thus $r/\Delta r \rightarrow 0$. On the other hand for comets that suffocate due to a thick superficial layer of dust, r is about constant and $\Delta r \rightarrow 0$. Thus $r/\Delta r \rightarrow \infty$. In other words, sublimating comets move down and suffocating comets move up. *This is one of the most important characteristic of this plot* (Figure 20).

As a consequence, the two vertical areas have to be separated by a border, above, suffocating-away comets and below, sublimating-away objects. The current estimate of the location of the border is at (4±3)x10^4 returns. This puts forward a complex question on the theoretical meaning of this border. Our own interpretation is that the border represents the limit in which the gas flux is no longer capable of lifting all the dust, and thus some dust falls back on the surface thickening the layer with time. Of course there is much more to it than this shallow analysis, but to find out is beyond the scope of this paper.

**7.2 The A'Hearn-Arpigny calibration for the dust**

In order to plot the location of 133P, 249P, 331P, 62412 and 6478 in the RR vs ML-Age diagram, we have to calculate the production rate of dust along the orbit and integrate it. We are not going to include gas because there is no evidence of gas in these comets (Chandler et al., 2019; Hui et al., 2019; Jewitt et al., 2019; Kleyna et al., 2019; Moreno et al., 2019; Ye et al., 2019). To do this we are going to adopt the definition of Afρ by A'Hearn et al. (1984, 1995). Afρ was adopted as a theoretical model of the dust to remove the aperture dependence from the photometry. The idea was to put the different apertures on a common footing. A'Hearn et al. (1984, 1995) found that for the majority of comets and for apertures of 10^3 to 10^5 km, the model works well and Afρ was reasonably constant with aperture. The Afρ value is computed from

$$A f\rho\ [cm] = [\ (2.R.\Delta)^2 / \rho\ ]\ 10^{0.4[m(Sun)-m(obs)]} \quad (12)$$

where R is the comet-Sun distance in AU, Δ is the Earth-comet distance in cm, ρ is the physical radius in cm of the photometric aperture at the distance of the comet, m(Sun) = -27.07 is the solar R-band magnitude, and m(obs) is the phase angle corrected R-band magnitude of the comet inside a 4"-radius aperture. These Afρ values have still to be corrected by phase effects.

Afρ comes expressed in cm. In order to be useful we need this quantity expressed in kg. On March 20th, 2009, we wrote to Mike A´Hearn and asked him what the conversion factor would be and this was his answer (quote): *"If I want an interpretation I use another empirical correlation that was worked out by Claude Arpigny (Arpigny et al., 1986), and which is discussed in our 1995 paper on the photometry database (A'Hearn et al., 1995). That relation as stated there is that an Afρ of 1000 cm is equivalent to a mass loss rate of one metric ton per second. Equivalently this means that an Afρ of 1 cm is equivalent to a mass loss rate of one kg per second. This relationship, of course, ignores differences in particle size distribution, differences in outflow velocity, and lots of other things, but it is useful in getting an understanding of patterns as long as it is not over-interpreted"* (private communication). Accordingly, we will call this the A'Hearn-Arpigny dust calibration.

A web site managed by Castellanos (2019), collects magnitude observations of comets and calculates the Afρ for many of them, including 133P, 249P and 6478. The last object has the best and longest data set and thus will be used as a reference.

The calibration Afρ (corrected for phase effects) vs m(1,1,α) is shown in Figure 16 for comet 6478 Gault. Since our five objects belong to a tight class that exhibit very similar morphological and physical properties (see for example Table 4), it is a reasonable assumption that this same calibration can be applied to all of them. This procedure has the advantage that the same measuring stick is applied to the objects. The representation on the RR vs ML-Age diagram sheds new light on the nature of these objects because it will compare them to other groups of comets.

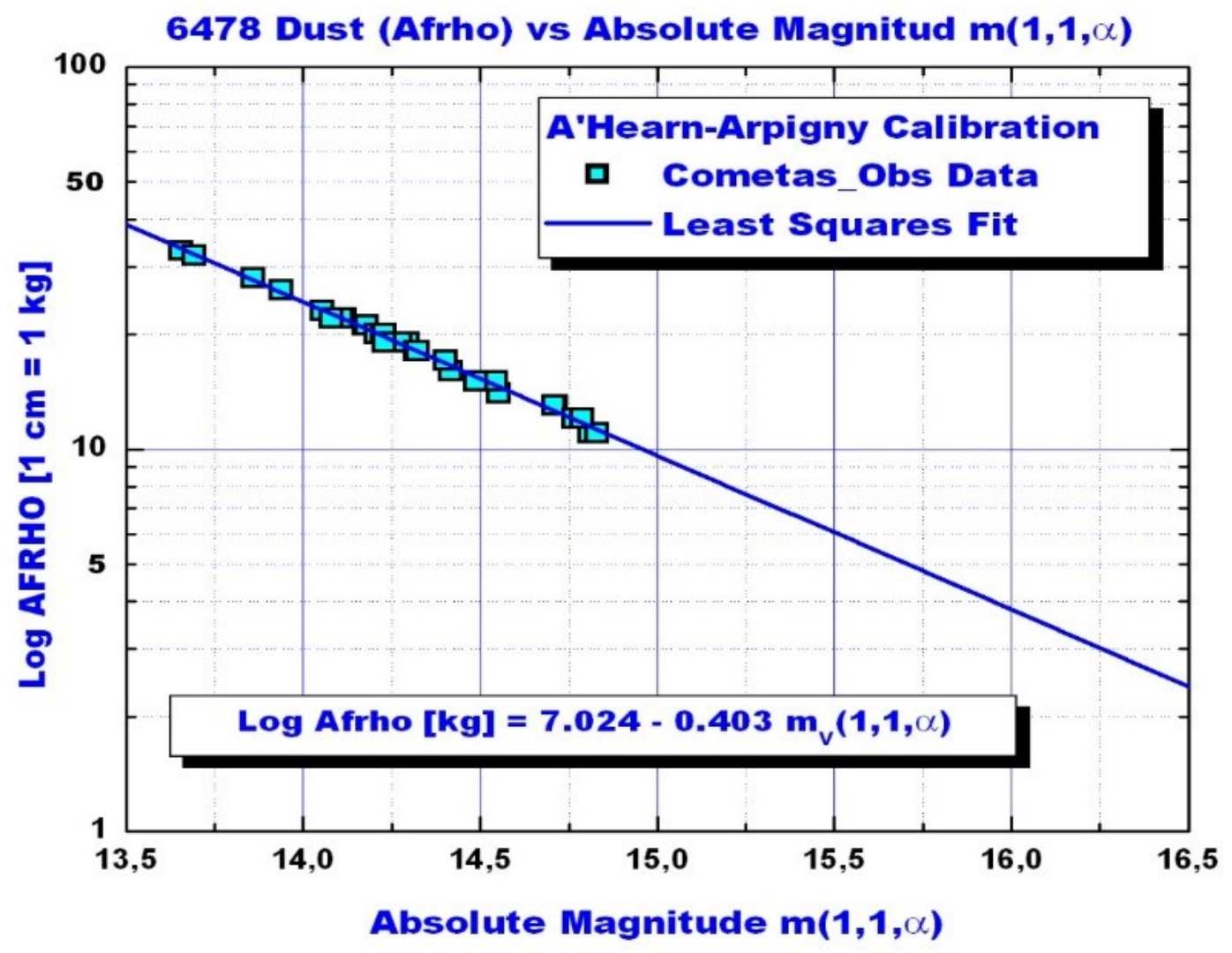

**Figure 19.** The calibration A'Hearn-Arpigny is based on their analysis (A'Hearn et al. 1984, 1995; Arpigny et al. 1986; and personal communication).

<> <> <> <> <>

### 7.3 Comet 133P: Remaining Returs, ML-Age and thickness lost per apparition

Using the minimum density found by Jewitt et al. (2004) ρ = 1300 kg/m3, we find a total mass for the nucleus $M_{NUC}$ = 5.1x10^13 kg. Using the A'Hearn-Arpigny´s calibration, we find a mass loss per apparition ML = 5.7x10^8 kg. Thus RR = r/Δr = 3 M/ΔM = 2.7x10^5 returns. Then we can calculate the depth of the layer lost per apparition, Δr = 7.8 mm per return. The ML-Age can be computed from equation (8), ML-Age = 628 cy. To make sense of these results it should be remembered that comet 28P/Neujmin 1 has an age of 100 cy. If those values are plotted in the RR vs ML-Age diagram (Figure 20) we find that 133P lies in the suffocation region above RR = 4x10^4 returns where r/Δ r (t) → ∞ , but outside the Graveyard region (ML-Age > 1000 cy).

### 7.4 Comet 249P/LINEAR: Remaining Returns, ML-Age, thickness lost per apparition

Above we have found the mass of 249P, M = 6.6x10^12kg. Using the A'Hearn-Arpigny´s calibration, we find a mass loss per apparition ML = 3.2x10^7 kg. Thus RR = r/Δr = 3 M/ΔM = 2.7x10^6 returns. Then we can calculate the depth of the layer lost per apparition, Δr = 0.8 mm per return. And the ML-Age can be computed from ML-Age = 3.58x10^11 / ML = 6280 cy. If the values are plotted in the RR vs ML-Age (Figure 20) we find that 249P lies in the suffocation region well above the border at RR = 3x10^5 returns, and inside the Graveyard. The dust thickness lost per apparition is 10 mm.

### 7.5 331/Gibbs: Remaining Returns, ML-Age, thickness lost per apparition

Above we have found a mass for 331P M = 3.8x10^14 kg. Applying the A'Hearn-Arpigny calibration we find a dust Budget ΔM = 2.0 x 10^8 kg. Then we find M/ΔM = 1.9x10^6 and RR = r/Δr = 5.7x10^6, from which the layer lost per apparition is Δr = 0.7 mm. From the equation (8) we obtain ML-Age = 1.8x10^3 cy. When we plot 331P on the RR vs ML-Age diagram (Figure 20) it lies on the suffocation region well above the border at RR = 4x10^4 returns, and inside the Graveyard.

### 7.6 62412: Remaining Returns, ML-Age, thickness lost per apparition

Since we have not been able to identify an active region in the SLC of this comet (Figure 17), we do not know about its activity. Thus it will not be possible to plot this object in our evolutionary diagram.

### 7.7 6478 Gault: Remaining Returns, ML-Age, thickness lost per apparition

To calculated the mass loss from this comet, we have to assume that the six regions are due to dust activity on the comet. If this is not the case our calculation would have to be revised. We may be able to solve this riddle on 2020 because the observing window covers the active Zone 5 region (Figure 3 of Ferrín et al., 2019b).

We found that 6478 is double with possible primary radius $r_P$ = 1850 m, 1550 m, 1390 m, and secondaries $r_S$ = 660 m, 552 m and 495 m. In order to get these results a minimum density of $\rho_{MIN}$ = 1590 kg/m$^3$ had to be adopted. Since the evolutionary diagram is a log-log plot with 9 orders of magnitude in the vertical axis and 6 orders of magnitude in the horizontal axis, the diagram is very forgiving. Thus to avoid plotting 3 data points for the same object, we are going to adopt the mid values for the primary and secondary. It will be found that the mass contributed by the secondary is negligible.

We find a total mass for the binary M(p+s) = 2.6x10^13 kg. Thus M/ΔM = 27000 and RR = r/Δr = 81000 returns. For the age we apply equation (8) and find ML-Age = 373 cy, even younger than 133P/Elst-Pizarro. And for the thickness of the layer lost per apparition we find 2 mm. When we plot this values on the evolutionary diagram (Figure 20) we find that 6478 plots in the suffocation region, but not yet entering the Graveyard. In fact 6478 is the youngest member of the PCG.

## 8. Agreement between theory and location in the plot

Based on the current theoretical interpretation (Chandler et al., 2019; Hui et al., 2019; Jewitt et al., 2019; Kleyna et al., 2019; Moreno et al., 2019; Ye et al., 2019), 6478 is in a regime of slow rotational disruption, shedding dust slowly and episodically due to its fast rotation. The fast rotation is an indication of an evolved object, and the shedding of dust is an indication of a superficial source of dust (which a sublimating comet will not have). The absence of gas means that the ice layer lies deeper than the thermal wave penetration. This explanation for 6478 must be equally valid for the other 4 members of the PCG.

In the evolutionary diagram (Figure 20) the region above RR=4x10^4 is the Dust Suffocation Region. The fact that 133P, 249P, 331P and 6478 are in the Rotational Disruption Regime agrees completely with the region of the evolutionary diagram in which they lie ($r/\Delta r \rightarrow \infty$), giving credence to two concepts: the comets are suffocating and this is an evolutionary stage at the end of their lifetime.

When the 4 objects for which we have data are plotted on the evolutionary diagram, they all lie above the upper limit of the border line RR = 7x10^4 returns. So there is an agreement between theory and observations. The diagram correctly interprets the evolutionary regime of these objects, and places them close by on the same general region of suffocation.

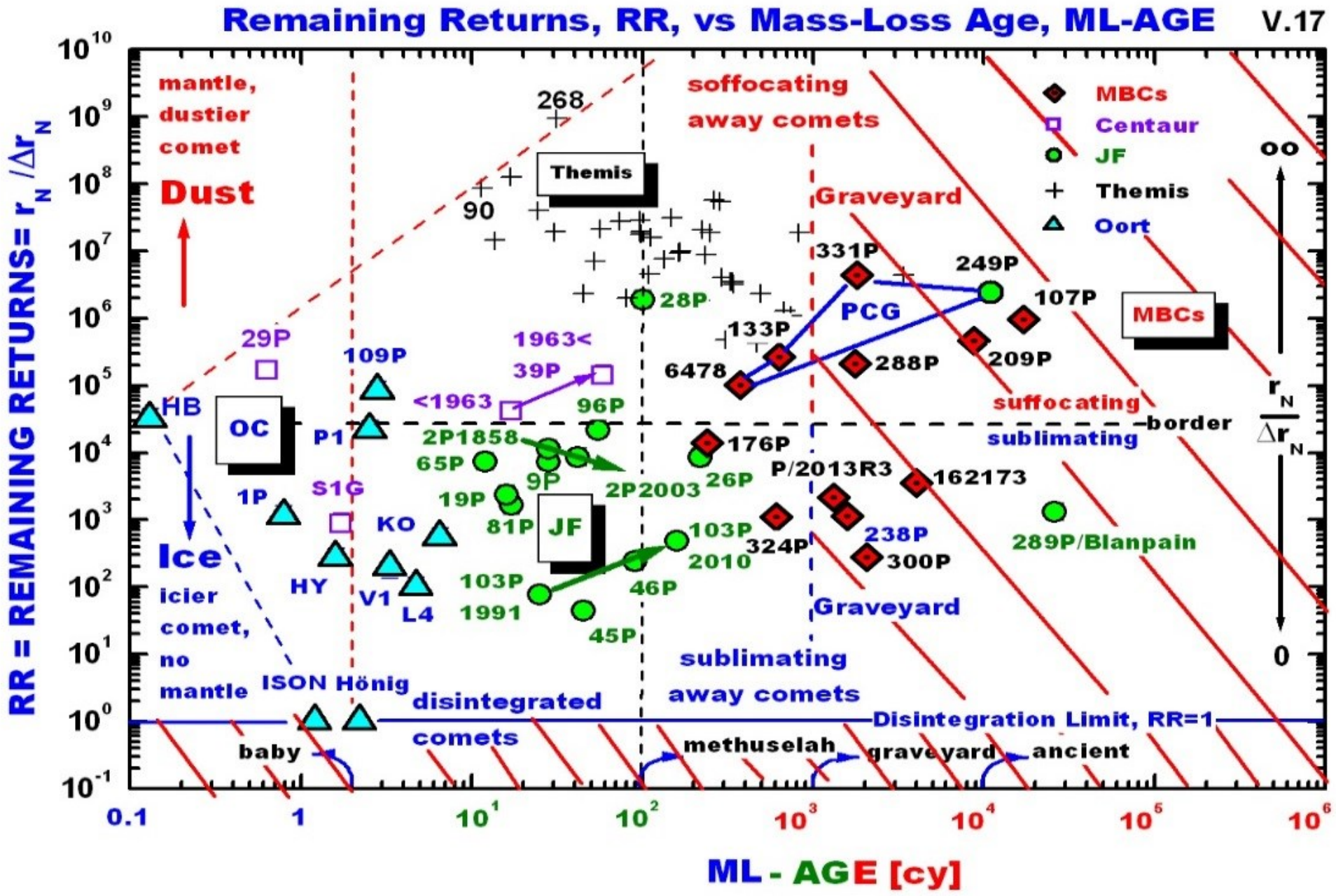

**Figure 20.** Remaining Returns vs Mass-Loss Age diagram. (1) The diagram covers a large area: nine orders of magnitudes on the vertical axis, six on the horizontal axis. Because the diagram is log–log, the diagram is very forgiving. (2) If a comet were made of pure ice, the layer removed by apparition, Δr, would remain approximately constant (see text), and r/Δr would tend to zero as the comet sublimates away. If the comet contained much dust, part of it would remain on the surface, Δr would tend to zero, and r/Δr would tend to infinity. Thus sublimating-away comets move down, and suffocating comets move up in the diagram. Necessarily there is a border that separates these two regions, sublimating (below) and suffocating (above). (3) The area with ML-AGE > 1000 cy is defined as the Graveyard. (4) The four members of the PCG are united by a polygon. 6478 and 133P, are not yet in the graveyard. 331P and 249P are in the graveyard. 6478 is the youngest of the group, 249P the oldest. (5) 249P is a member of the Jupiter Family. The other three are members of the MBCs. Comet 62412 could not be plotted because the area of activity has not been identified. (6) The diagram separates comets into classes: Oort Cloud comets on the left, Jupiter Family comets in the middle, MBCs on the right, and disintegrating comets below on the RR = 1 line. (7) It is not surprising that MBCs occupy the right-hand side of the diagram: this is expected on physical grounds. The diagram shows that they are old (large ML-Age) and that they have a substantial dust or crust layer (large r/Δr). (8) Two comets are plotted on the disintegration limit: C/2002 O4 Hönig and ISON. (9) Objects move on the diagram. Comet 39P had a close encounter with Jupiter in 1963 and the orbit changed substantially. Comet 2P/Encke shows evolution of the parameters between the 1858 and the 2003 apparitions. Comet 103P shows evolution between the 1991 and the 2010–2011 apparitions. Motion is also expected for the PCG but has not been detected yet. (10) Much theoretical and observational work remains to be done to elucidate the full meaning and the limitations of the RR vs ML-Age diagram. (Ferrín et al. 2012). Nomenclature: 1P, 1P/Halley; KO, C/1973 E1 Kohoutek; P1, C/2009 P1 Garradd; V1, C/2002 V1 NEAT; HB, C/1995 O1 Hale–Bopp; L4 = C/2011 L4 Panstarrs; HY, C/1996 B1 Hyakutake; Ho, C/2002 O4 Hönig; S1G, P/2009 S1 Gibbs. MBC, Main Belt Comet; OC, Oort Cloud; JF, Jupiter Family. PCG = Pinpoint Comet Group.

## 9. Conclusions

(1) 133P/Elst-Pizarro. The SLC presented in this work is one of the best SLC we have been able to compile. The turn on and turn off points are clearly defined. We found $m_V(1,1,0) = 15.83\pm0.05$ and a radius $r = 2113\pm400$ m. The LAG from perihelion is ~170 d suggesting that the pole of this object might be tilted toward the Sun at this place into the orbit. We show evidence for the existence of a tail in the light curve in the range $210\text{ d} < t - Tq < 550\text{ d}$, that might be due to CO or CO2 sublimation.

(2) 249P/LINEAR. This is a Jupiter Family comet, implying that the condition of PCG is not restricted to main belt objects. We found $m_V(1,1,0) = 16.13\pm0.05$. The object is double with equal size components. Assuming geometric albedos of 0.042, 0.06, 0.075 we find radii 1378 m, 1153 m, 1031 m. This comet was a PCG member on 2006, but due to a perihelion distance diminution, its insolation is increasing and on 2015-16 left the group and became a bona fide comet with gas coma and tail. However this condition is not going to last for long. On 2019 the comet will receive less energy from the Sun than on 2006, presumably returning to the PCG. Because of the perihelion distance diminution, this is a good example of a Lazarus comet.

(3) 331P/Gibbs. There is very little information on this object and the SLC is poorly determined. Even so it was possible to measure most of the relevant parameters. We found $m_V(1,1,0) = 14.50\pm0.05$, and a radius of $r = 4125\pm350$ m.

(4) 62412. We acquired photometric data for this object and created the SLC which is uneventful and with no activity. The previously detected activity lies far from the Sun but was beyond our observing window. There is only one single paper (and image) showing the activity of this object, but was included in this category because its rotational period is Prot = 3.33 hours (Sheppard and Trujillo, 2010). We found $m_V(1,1,0) = 13.90\pm0.05$, and a radius of $r = 5390\pm610$ m.

(5) 6478 Gault. The phase, SLC and rotational light curve of this object have studied in previous papers ( Ferrin and Acosta, 2019a; Ferrín et al., 2019b). The object exhibits 6 zones of activity which complicate the understanding of its secular activity. We determined the rotational period Prot(2)= 3.360±0.005 hours exhibiting a rotational light curve with two peaks and amplitude Amp = 0.16±0.02 magnitudes. This value is similar to the rotational periods of other comets (133P and 62412). We determined $m_V(1,1,\alpha)= 16.12\pm0.05$. The system is double with unequal components. We determined the radius of the two components for values of the geometric albedo 0.042, 0.06 and 0.075 and found $r_P$ = 1850 m, 1550 m, 1390 m, and for the secondary, $r_S$ = 660 m, 550 m, 495 m.

(6) We established the calibration A´Hearn-Arpigny for the dust production as a correlation between Afρ [kg] vs $m_V(1,1,0)$. Notice that now Afρ is given in kg.

(7) We calculated the quantities $r/\Delta r$ = RR = Remaining Returns and ML-Age, a proxy for age, for 4 of the 5 objects we are studying, and plotted their location on an evolutionary diagram (Figure 20). It was not surprising that the 4 objects lie nearby to each other, and in the correct region of the diagram. The objects are suffocating and getting rid of their suffocation by means of centrifugal forces. The PCG objects are in a slow rotational disruption regime.

(8) Our rotational period for 6478 Gault was confirmed by Carbognani and Buzzoni (2019) who found a rotational period Prot = 3.358±0.005 hours in excellent agreement with our result. However in their light curve the eclipse is no longer visible, leaving the light curve with one single minimum per rotation which is interpreted as an albedo feature.

(9) We found a border separating two distinct regions on the RR vs ML-Age diagram, whose current estimated location is (4±3)x10^4 returns. Although our preliminary interpretation is that this is the regime where the gas intensity is no longer capable of lifting all the dust, the full meaning of this border has still to be grasped and this task is beyond the scope of this paper.

(10) We can now to attempt to answer the questions we raised at the beginning: What do these objects have in common? Basically their morphology, but the fact that they are in a slow rotational disruption regime implies that their rotational periods are near the rotational limit, and three comet for which we have rotational data have a mean rotational period <Prot> = 3.39±0.05 h, a very sharp distribution. The are located nearby in the evolutionary diagram (Figure 20). In what do they differ? Their secular light curves are different although all are of low intensity. Are they sublimating or suffocating? The evidence points to suffocation as the regime of these objects. Interestingly they are getting rid of their suffocation by means of centrifugal forces. Is the PCG restricted to Main Belt of comets? No. 249P is a Jupiter family comet. This is telling us that this evolutionary condition is not restricted to the main belt. Do they represent a terminal stage in their lifetime? We think so. Some objects evolve into a rotational disruption stage while other do not. Which evolve and how evolve into this phase is to be determined. Why did 249P left the group after 2006? Because of a diminution of its perihelion distance that increased the energy coming from the Sun, classifying the object as a Lazarus comet. Is there evidence of duplicity in any of them? Yes, 2 of 5 objects exhibit evidence of duplicity, 249P and 6478/Gault. May their density be considered normal? From Table 4 we see that the density $\rho > 1300$ kg/m3 is much larger than the density of Jupiter Family comets $\rho < 600$ kg/m3 found by Kokotanekova et al. (2017).

## 10. Acknowledgements

The FACom group is supported by the project 'Estrategia de Sostenibilidad 2015–2016', sponsored by the Vicerectoría de Investigation of the Universidad de Antioquia, Medellín, Colombia. This work contains observations made at the National Observatory of Venezuela, ONV, Centro de Investigaciones de Astronomía, CIDA, Mérida. The help of Giuliatna Navas and the night assistants The FACom group is supported by the project 'Estrategia de Sostenibilidad 2015–2016', sponsored by the Vicerectoría de Investigation of the Universidad de Antioquia, Medellín, Colombia. Richard Rojas, Leandro Araque, Dalbare Gonzalez, Fredy Moreno, Carlos Pérez, Gregory Rojas, Daniel Cardozo and Ubaldo Sánchez, is particularly appreciated. We used as comparison stars the APASS catalogue published by the AAVSO. We acknowledge with thanks the comet observations from the COBS Comet Observation Database, contributed by observers worldwide and used in this research. The contribution of images by Jesús Canive, Edwin Quintero and José Francisco Hernández is appreciated. Our thanks go to Julio Castellanos in charge of the http://www.astrosurf.com/cometas-obs/ web site, for this sharing of data.

TABLES

Table 1. Instrumentation

| Observatory | MPC Code | City | Country | Diameter | CCD | Observer |
|---|---|---|---|---|---|---|
| Galileo Galilei<br>Lat-31°49'22.96"<br>Lon60° 31' 14.3" | X31 | Oro Verde<br>Entre Rios | Argentina | 35 cm | Sbig<br>8300M | Cesar<br>Fornari |
| | Z39 | Lanzarote<br>Canary<br>Islands | Spain | 24 cm | ST-8XME | Agustin<br>Acosta |
| CIDA<br>Lat 8° 47'<br>Lon 70° 52' W<br>Alt 3590 m | 303 | Mérida | Venezuela | 100 cm | Techtronic | Ignacio<br>Ferrín |

Table 2. Observing log of 62412

| YYYYMMDD | $m_V(1,1,0)$ | t–Tq [d] | Observer | #images |
|---|---|---|---|---|
| 171209 | 14.05±0.02 | -324 | Ferrín | 11 |
| 171223 | 13.93±0.01 | -310 | Ferrín | 12 |
| 190225 | 13.83±0.01 | 119 | Acosta | 6 |
| 190311 | 13.94±0.01 | 134 | Acosta | 22 |
| 190429 | 13.89±0.02 | 183 | Acosta | 28 |
| 190511 | 13.83±0.01 | 195 | Fornari | 35 |
| 190527 | 14.09±0.01 | 211 | Fornari | 6 |
| 190608 | 14.21±0.02 | 224 | Fornari | 19 |

Table 3. Physical parameters of the PCG.

| object | $m_V(1,1,0)$ | D [km]<br>r [m] | Asec(q) | Δt active [d] | Δr [mm] | RR=r/Δr [returns] | ML-Age [cy] | β [mag/°] |
|---|---|---|---|---|---|---|---|---|
| 133P | 15.83±0.05 | 4.2±0.8<br>2110 | -2.4±0.01 | 170 | 9 | 2.7x10^5 | 628 | no phase effect |
| 249P<br>S1<br>S2 | 16.12±0.05 | 3.89<br>1950 | -----<br>-0.75<br>-0.27 | -----<br>48<br>27 | 0.05 | 3.8x10^6 | 11000 | 0.034 |
| 331P | 14.50±0.05 | 8.25<br>4120 | -0.83 | 60 | 10 | 40000 | 1800 | 0.0328 |
| 62412 | 13.90±0.05 | 10.7<br>5390 | no data | no data | no data | no data | no data | 0.0488 |
| 6478 | | | | | | 81000 | 373 | no phase effect |

Table 4. Rotational Periods

| Objects | Rotational Period [h] | ρ [gm/cm$^3$] | Reference | classification |
|---|---|---|---|---|
| 133P | 3.471 | 1.3< | Hsieh et al. (2004) | MBC, Themis Methuselah, Lazarus |
| 249P | no data | no data | ---- | Jupiter Family |
| 331P | no data | no data | ---- | MBC |
| 6478 | 3.360±0.005 | 1.59≤ | This Work | MBC, Phocaea |
| 62412 | 3.33 | ~1.50 | Sheppard and Trujillo (2014) | MBC, Methuselah, Lazarus |

Table 5. Orbital Elements and Tisserand Parameter of the Pinpoint comets.
Asteroids have $T_J > 3.0$; Comets of the Jupiter Family have $2.0 < T_J < 3.0$

| OBJECT/ FAMILY | q[A] | Q[AU] | e | i[°] | a[AU] | Tq yyyymmdd | Porbit [y] | Tiss |
|---|---|---|---|---|---|---|---|---|
| 133P/ Elst-Pizarro | 2.66 | 3.66 | 0.16 | 1.39 | 3.16 | 20180921 | 5.62 | 3.2 |
| 249P/ LINEAR | 0.50 | 5.027 | 0.82 | 8.40 | 2.76 | 20200629 | 4.59 | 2.7 |
| 331/Gibbs | 2.88 | 3.13 | 0.04 | 9.74 | 3.00 | 20200930 | 5.21 | 3.2 |
| 6478 Gault | 1.86 | 2.75 | 0.19 | 22.8 | 2.30 | 20200103 | 3.5 | 3.5 |
| 62412 | 2.90 | 3.41 | 0.08 | 4.73 | 3.15 | 20181029 | 5.59 | 3.2 |